\documentclass[aps,prd,twocolumn,superscriptaddress,nofootinbib,floatfix]{revtex4-2}

\usepackage{amsmath,amssymb}
\usepackage{graphicx}
\usepackage{bm}

\newcommand{\fR}{f(R)}
\newcommand{\fRz}{|f_{R0}|}
\newcommand{\Geff}{G_{\rm eff}}
\newcommand{\Mpc}{h^{-1}{\rm Mpc}}
\newcommand{\lcdm}{\Lambda{\rm CDM}}

\newcommand{\xivg}{\xi_{vg}}

\begin{document}

\title{Unveiling $f(R)$ Gravity with Void-Galaxy Cross-Correlation Multipoles}

\author{Yue Nan}
\email{nan@tokuyama.ac.jp}
\affiliation{Department of Mechanical and Electrical Engineering,\\ National Institute of Technology, Tokuyama College,\\Gakuendai, Shunan, Yamaguchi 745-8585, Japan}
\affiliation{Kavli Institute for the Physics and Mathematics of the Universe (WPI), The University of Tokyo Institutes for Advanced Study (UTIAS), The University of Tokyo, Kashiwa, Chiba 277-8583, Japan}

\begin{abstract}
Cosmic voids provide low-density environments where the scalar fifth force
predicted by $f(R)$ modified gravity can be weakly screened.
We present a semi-analytical calculation of the monopole, dipole, and
quadrupole of the void-galaxy cross-correlation function $\xi^{s}(s,\mu)$ in
redshift space for the Hu-Sawicki $f(R)$ model ($n=1$), combining
scale-dependent growth induced by the scalaron with nonlinear spherical shell
dynamics.  The same framework can be generalized to metric $f(R)$ theories for
which $G_{\rm eff}(k,a)/G$ is specified in the quasi-static limit.
Our key results are:
(1)~the monopole deviation from $\Lambda{\rm CDM}$ grows from $+2.8\%$ for
large voids ($r_v=30 h^{-1}{\rm Mpc}$) to $+29.7\%$ for small voids
($r_v=11.7 h^{-1}{\rm Mpc}$) at $|f_{R0}|=10^{-5}$, a distinctive
size-dependent signature of the Compton-scale scalaron response, with
$\lambda_C\approx 8 h^{-1}{\rm Mpc}$;
(2)~nonlinear evolution amplifies the modified-gravity signal by
$\mathcal{A}_0\approx 4$, bringing it within reach of ongoing and upcoming
spectroscopic surveys such as DESI, Subaru PFS, Euclid, and Roman;
(3) the gravitational potential contains a finite-range Yukawa
component, producing a radially dependent dipole signature complementary to the
density and velocity multipoles;
(4) for the fiducial Hu-Sawicki evolution, the signal generally decreases
toward higher redshift as the scalaron Compton wavelength becomes shorter,
but remains potentially detectable at Stage-IV spectroscopic void samples.
We show that the void-scale transition in the modified-gravity response, 
the joint sensitivity to density, velocity, and fifth-force contributions, 
and the nonlinear amplification around void shells make redshift-space 
void-galaxy multipoles a powerful semi-analytical probe of $f(R)$ gravity 
and effective dark-energy inhomogeneities in modified gravity.

\end{abstract}

\maketitle
\vspace{-18pt}

\section{Introduction}
\label{sec:intro}

The accelerating expansion of the Universe, established through
Type Ia supernovae observations~\cite{Riess:1998cb,Perlmutter:1998np},
may be explained by a cosmological constant within general relativity
(GR), but the theoretical challenges associated with the vacuum 
energy~\cite{Weinberg:1988cp} motivate the exploration of alternative
explanations.  One well-studied class of alternatives to GR is $\fR$
modified gravity, in which the Einstein-Hilbert action is supplemented
by a general function of the Ricci scalar
$R$~\cite{Sotiriou:2008rp,DeFelice:2010aj,Clifton:2011jh,Nojiri:2010wj,Nojiri:2017ncd}.  Among
the various $\fR$ models, the Hu-Sawicki (HS) model~\cite{Hu:2007nk}
is particularly attractive because it is designed to pass Solar System
tests of gravity through the chameleon screening
mechanism~\cite{Khoury:2003aq,Jain:2010ka}; other viable $\fR$
Lagrangians that unify inflation with late-time acceleration while
satisfying local gravity tests have also been
constructed~\cite{Nojiri:2007as,Nojiri:2007cq,Cognola:2007zu}.

Cosmic voids---the large underdense regions of the cosmic web with
characteristic scales of order $\mathcal{O}(10)\,h^{-1}{\rm Mpc}$
\cite{Pisani:2019cvo,Sutter:2012tf,Sheth:2003py}---provide a
promising arena for testing modified gravity, including $f(R)$ gravity.
Because the chameleon mechanism is less efficient in regions with
shallow gravitational potentials, void interiors and their surrounding
low-density environments can be partially unscreened.  In these regions,
the additional scalar degree of freedom, the scalaron, can mediate a
fifth force and enhance the effective gravitational interaction
\cite{Clampitt:2012tm,Cai:2014efa,Falck:2017rvl,Tamosiunas:2022csc}.  The void-galaxy
cross-correlation function and its redshift-space distortion (RSD)
multipoles encode complementary information about the density profile
and the coherent velocity field around voids
\cite{Hamaus:2015gua,Hamaus:2016sei,Cai:2016jek,Nadathur:2019mct}.
Moreover, odd multipoles of the void-galaxy cross-correlation can
receive relativistic light-cone contributions, including gravitational
redshift and potential-gradient effects, making them sensitive to the
gravitational potential and hence to modifications of the Poisson
equation, such as the scale- and environment-dependent enhancement of
$G_{\rm eff}/G$ in $f(R)$ gravity \cite{Nan:2018tce}.

Several groups have studied void statistics in $\fR$ gravity using
$N$-body simulations~\cite{Li:2011pj,Cautun:2018gae,Paillas:2018wxs}
and have placed observational constraints on $\fRz$ from void
abundance~\cite{Achitouv:2016mbn,Contarini:2021fkv,Contarini:2023qqx},
void lensing~\cite{Baker:2019gxo,Davies:2019irs}, and the void-galaxy
cross-correlation~\cite{Hamaus:2021lzy,Woodfinden:2022bhx}.
Independent constraints from galaxy clusters and
clustering~\cite{Cataneo:2014kaa,Terukina:2013eqa,Wilcox:2015kna,Mitchell:2021uzh},
marked correlation functions~\cite{Armijo:2024mga,Armijo:2024mgb},
galaxy-scale tests~\cite{Desmond:2020gzn},
the full-shape analysis of baryon acoustic oscillation
surveys~\cite{Alam:2020jdv,DESI:2024mwx}, and cluster mass
functions~\cite{Hagstotz:2018onp} currently bound
$\fRz \lesssim 10^{-5}$ to $10^{-6}$.

In this work, we extend the analytical framework developed in
Nan \& Yamamoto~\cite{Nan:2018tce} to predict the monopole, dipole,
and quadrupole of the void-galaxy cross-correlation in Hu--Sawicki
$f(R)$ gravity.  Our extension incorporates the scale- and
time-dependent growth induced by the modified Poisson equation and
propagates it into the Fourier-space evolution of the void density
profile.  This enables us to quantify how the scalaron-mediated fifth
force modifies the density, velocity, and potential fields around
voids.  We focus on the void-size dependence of the signal, a
characteristic imprint of chameleon screening, and on the enhancement
of the predicted multipoles generated by nonlinear spherical shell
evolution.  Finally, we present estimations for the
detectability of the $f(R)$ signal in void-galaxy cross-correlation
multipoles for ongoing and upcoming spectroscopic surveys.

The paper is organized as follows.
Section~\ref{sec:fR} presents the linear theoretical framework: the
Hu-Sawicki $\fR$ model, the scale-dependent growth, the void density
profile, the velocity and gravitational-potential fields, and the RSD
multipole formulae for the void-galaxy cross-correlation.
Section~\ref{sec:nonlinear} describes nonlinear spherical void evolution
and the resulting MG signal amplification.
Section~\ref{sec:results} presents the fiducial numerical setup,
results, and detectability estimates for the $f(R)$ signal.
We conclude in Sec.~\ref{sec:conclusions}.
Throughout we adopt natural units with $c = 1$ except where stated otherwise.

\section{Theoretical Framework of Linear $f(R)$ Gravity with Void RSDs}
\label{sec:fR}

\subsection{Modeling $f(R)$ Gravity Effects on Structure Growth}

\subsubsection{Action of $f(R)$ gravity and the Hu-Sawicki model}

In $\fR$ gravity the gravitational action is
\begin{equation}
\label{eq:action}
S = \int d^4 x \sqrt{-g}\,\frac{R + f(R)}{16\pi G} + S_{\rm m}\,,
\end{equation}
where $S_{\rm m}$ is the matter action.  Variation with respect to the
metric yields a fourth-order equation that can be recast as a
second-order system by introducing the scalar field $f_R \equiv
df/dR$, the scalaron~\cite{Sotiriou:2008rp,DeFelice:2010aj}.

The HS model~\cite{Hu:2007nk} takes the functional form
\begin{equation}
\label{eq:HS}
f(R) = -m^2\,\frac{c_1\,(R/m^2)^n}{c_2\,(R/m^2)^n + 1}\,,
\end{equation}
where $m^2 \equiv H_0^2\,\Omega_{\rm m0}$ sets the mass scale.  The
ratio $c_1/c_2$ is fixed by requiring the same expansion history as
$\lcdm$:
\begin{equation}
\frac{c_1}{c_2} = 6\,\frac{\Omega_\Lambda}{\Omega_{\rm m0}}\,.
\end{equation}
For $n = 1$ the model has a single free parameter, the present-day
background value of the scalaron field,
\begin{equation}
f_{R0} \equiv \left.\frac{df}{dR}\right|_{z=0}
  = -\frac{c_1}{c_2^2}\,\left(\frac{m^2}{R_0}\right)^2\,,
\end{equation}
where $R_0$ is the background Ricci scalar today.  For general HS
index $n$, the high-curvature limit gives
$|f_R(a)|=|f_{R0}|(R_0/R_a)^{n+1}$, so changing $n$ changes the
redshift evolution of the scalaron mass even at fixed $|f_{R0}|$.

\subsubsection{Scalaron mass and effective gravitational constant}

For a general metric $f(R)$ theory written as $R+f(R)$, we define
\[
        f_R \equiv {df\over dR},
        \qquad
        f_{RR}\equiv {d^2f\over dR^2}.
\]
The scalaron mass evaluated on the cosmological background is
\begin{equation}
        m_{\rm sc}^2(a)
        =
        {1\over 3}
        \left[
        {1+f_R(a)\over f_{RR}(a)}
        -
        R_a
        \right],
\end{equation}
where \(R_a\) denotes the background Ricci scalar at scale factor \(a\).
In viable high-curvature models,
\[
        |f_R(a)|\ll 1,
        \qquad
        {1\over f_{RR}(a)} \gg R_a ,
\]
so that the scalaron mass reduces to
\begin{equation}
        m_{\rm sc}^2(a)
        \simeq
        {1\over 3f_{RR}(a)} .
\end{equation}
Thus, once the background expansion is specified, the linear response of
a viable metric \(f(R)\) model in the high-curvature regime is controlled
by \(f_{RR}(a)\), or equivalently by \(m_{\rm sc}(a)\).

For the Hu--Sawicki model, the high-curvature expansion gives
\begin{equation}
        f_R(a)
        =
        f_{R0}
        \left({R_0\over R_a}\right)^{n+1},
        \qquad f_{R0}<0 .
\end{equation}
Equivalently,
\begin{equation}
        |f_R(a)|
        =
        |f_{R0}|
        \left({R_0\over R_a}\right)^{n+1}.
\end{equation}
Differentiating the high-curvature expression with respect to \(R\)
gives
\begin{equation}
        f_{RR}(a)
        \simeq
        -{n+1\over R_a} f_R(a)
        =
        {n+1\over R_a}|f_R(a)| .
\end{equation}
Substituting this into the high-curvature scalaron mass yields
\begin{equation}
        m_{\rm sc}^2(a;n)
        =
        {R_a\over 3(n+1)|f_R(a)|}
        =
        {R_a\over 3(n+1)|f_{R0}|}
        \left({R_a\over R_0}\right)^{n+1}.
\end{equation}
For the fiducial \(n=1\) case, this becomes
\begin{equation}
        m_{\rm sc}^2(a)
        =
        {R_a\over 6|f_{R0}|}
        \left({R_a\over R_0}\right)^2 .
\end{equation}
The connection between the scalaron mass and the scale-dependent
gravitational response can be seen from the linearized scalaron
constraint.  In the sub-horizon quasi-static limit, the time derivatives
of the perturbations are subdominant compared with their spatial
gradients, and the scalaron perturbation obeys the Yukawa-type equation
\begin{equation}
        \left({k^2\over a^2}+m_{\rm sc}^2(a)\right)
        \delta f_R(k,a)
        \simeq
        {8\pi G\over 3}\,\delta\rho_{\rm m}(k,a),
        \label{eq:scalaron_constraint}
\end{equation}
up to the overall sign set by the Fourier and metric-potential
conventions.  Here \(k\) is the comoving wavenumber and
\(k_{\rm C}(a)\equiv a m_{\rm sc}(a)\) is the corresponding comoving
Compton scale.  The ratio \(k/(a m_{\rm sc})=k/k_{\rm C}\) therefore
determines whether a mode lies above or below the scalaron's Compton
scale. Eq.~\eqref{eq:scalaron_constraint} shows that the scalaron response is
suppressed for \(k/a\ll m_{\rm sc}\), while it becomes efficient for
\(k/a\gg m_{\rm sc}\).

Eliminating \(\delta f_R\) from the linearized metric and scalaron
field equations gives the modified Poisson equation
\begin{equation}
        {k^2\over a^2}\Psi(k,a)
        =
        -4\pi G_{\rm eff} \delta\rho_{\rm m}(k,a),
        \label{eq:modified_poisson_mu}
\end{equation}
where \(\Psi\) is the Newtonian potential and
\(G_{\rm eff}=\mu_{f(R)}(k,a)G\) parameterizes the scale-dependent effective
gravitational coupling felt by nonrelativistic matter:
\begin{equation}
        \mu_{f(R)}(k,a)
        \equiv
        {G_{\rm eff}(k,a)\over G}
        =
        {1\over 1+f_R(a)}
        {1+\dfrac{4}{3}\dfrac{k^2}{a^2m_{\rm sc}^2(a)}
        \over
        1+\dfrac{k^2}{a^2m_{\rm sc}^2(a)}} .
\end{equation}
For \(|f_R(a)|\ll1\), this reduces to
\begin{equation}
        \mu_{f(R)}(k,a)
        \simeq
        1+
        {1\over 3}
        {k^2\over k^2+a^2m_{\rm sc}^2(a)} .
        \label{eq:mufR}
\end{equation}
corresponding to the (reduced) comoving Compton wavelength of the scalaron,
$\lambda_C = 1/(a m_{\rm sc})$.
This form also provides the natural way to test other viable
$f(R)$ models: replace $m_{\rm sc}(a)$ by that implied by the candidate
Lagrangian or phenomenological mass history (for example by changing
the Hu--Sawicki family from the baseline $n=1$ to $n=2$, or by
adopting a Starobinsky-like or exponential $f(R)$ model), recompute
the scale-dependent growth, and then use
the resulting density, velocity, and potential profiles in the same
semi-analytical RSD formulae.
Eq.~\eqref{eq:mufR} relies on the quasi-static approximation,
which assumes $|\ddot{f}_R| \ll |k^2 f_R / a^2|$.  For the HS model
with $\fRz = 10^{-5}$, the scalaron oscillation frequency satisfies
$m_{\rm sc}/H \approx 400$--$500$ at $0 < z < 1$, so the quasi-static limit
is well justified with corrections of order $H^2/m_{\rm sc}^2 \sim
10^{-5}$--$10^{-6}$.

It is important to distinguish Eq.~\eqref{eq:mufR} from a full
nonlinear chameleon calculation.  In this paper we solve the
linearized scalaron response around the background density and use
the associated Compton scale as an effective screened-to-unscreened
transition scale for void profiles.  We therefore use the terms
``screened'' and ``unscreened'' in this Compton-response sense.  A
fully environment-dependent prediction would require solving the
nonlinear scalaron equation in the actual void, wall, and filament
environment; this is part of the simulation-calibration program
discussed in Sec.~\ref{sec:conclusions}.

\subsubsection{Scale-Dependent Linear Growth Factor}
\label{sec:growth}

Under the same sub-horizon quasi-static conditions used in
Eq.~\eqref{eq:mufR}, the linear matter density contrast
$\delta(\bm{k},a)=D(k,a)\delta_0(\bm{k})$ obeys the standard
modified-growth equation~\cite{Hu:2007nk,Song:2006ej,DeFelice:2010aj}
\begin{equation}
\label{eq:growth_ode}
D'' + \left(2 + \frac{d\ln H}{d\ln a}\right) D'
  = \frac{3}{2}\,\Omega_{\rm m}(a)\,\frac{\Geff(k,a)}{G}\,D\,,
\end{equation}
where primes denote derivatives with respect to $\ln a$.  The equation
assumes nonrelativistic matter, negligible radiation and anisotropic
stress for the late-time modes of interest, and $k/a\gg H$.  Due to
the $k$-dependence of $\Geff$, the growth factor $D(k,a)$ is itself
scale-dependent in $\fR$ gravity.

We define the growth rate
\begin{equation}
\label{eq:fka}
f(k,a) \equiv \frac{d\ln D(k,a)}{d\ln a}\,,
\end{equation}
and the growth ratio
\begin{equation}
\label{eq:Rka}
\mathcal{R}(k,a) \equiv \frac{D_{f(R)}(k,a)}{D_{\rm GR}(a)}\,.
\end{equation}
For $\fRz = 10^{-5}$ and $k \gg a m_{\rm sc}$, the ratio $\mathcal{R}$
exceeds unity, reflecting the enhanced growth from the fifth force.
At low $k$ ($k \ll a m_{\rm sc}$), $\mathcal{R} \to 1$ and GR is recovered.

Eq.~\eqref{eq:growth_ode} is integrated numerically from
$a_{\rm init} = 0.02$ ($z_{\rm init} = 49$) with initial conditions
set deep in the matter-dominated era where $\fR$ effects are
negligible.  At the linear level, the growth ratio $\mathcal{R}(k,a)$
is the exact quasi-static solution of the linearized scalaron
equation; we have verified internal consistency by reproducing the
same $\mathcal{R}(k)$ from an independent ODE integration to machine
precision.  The growth ratio transitions from $\mathcal{R} \approx 1$
at $k \ll a m_{\rm sc}$ to $\mathcal{R} \approx 1.14$ at $k = 1\;h/$Mpc for
$\fRz = 10^{-5}$ at $z = 0.5$, consistent with the Compton scale
$k_C \approx 0.12\;h/$Mpc.
At the \emph{nonlinear} level, quantitative comparison against
simulation-calibrated fitting
functions~\cite{Winther:2015wla} and full $N$-body void profiles is
deferred to future work (Sec.~\ref{sec:conclusions}).

\subsection{Void Density Profile}
\label{sec:profile}

\subsubsection{Universal profile}

We adopt the universal void density profile proposed by
Hamaus~\textit{et~al.}~\cite{Hamaus:2014afa}, which provides a
four-parameter analytic description of stacked void profiles measured
in $N$-body simulations:
\begin{equation}
\label{eq:delta_r}
\delta(r) = \Delta_c\,
  \frac{1 - (r/r_s)^\alpha}{1 + (r/r_v)^\beta}\,.
\end{equation}
Here $\Delta_c$ is the central underdensity ($\Delta_c < 0$ for
voids), $r_v$ is the effective void radius, $r_s$ is the compensation
scale beyond which the profile turns positive (the ``ridge'' around
the void), and $\alpha$, $\beta$ control the steepness of the inner
profile and the sharpness of the void wall, respectively.  This
functional form was shown~\cite{Hamaus:2014afa} to fit the mean
void-galaxy cross-correlation for a wide range of void sizes in
cosmological simulations, and has since been adopted as the standard
template for analytical void-RSD
studies~\cite{Hamaus:2015gua,Hamaus:2016sei,Cai:2016jek}.

We consider three representative void size classes
(Table~\ref{tab:void_params}).  The profile parameters are taken from
the void catalog analysis of Nan \&
Yamamoto~\cite{Nan:2018tce}, who fitted the universal
profile Eq.~\eqref{eq:delta_r} to stacked voids identified in the
SDSS/BOSS galaxy distribution.  The three size bins span the range
from large voids ($r_v = 30\;\Mpc$, for which $r_v \gg \lambda_C$
and the $\fR$ signal is small) to small voids ($r_v = 11.7\;\Mpc$,
where $r_v \sim \lambda_C$ and chameleon unscreening is maximal).

\begin{table}[t]
\caption{Void profile parameters from the universal fitting
function of Hamaus~\textit{et~al.}~\cite{Hamaus:2014afa}, with
numerical values from Table~II of Nan \&
Yamamoto~\cite{Nan:2018tce} (best-fit to stacked voids from
Ref.~\cite{Hamaus:2014afa}).  All radii in $\Mpc$; the scale
radius $r_s = (r_s/r_v)\times r_v$.}
\label{tab:void_params}
\begin{ruledtabular}
\begin{tabular}{lccccc}
Class & $r_v$ & $r_s/r_v$ & $\alpha$ & $\beta$ & $\Delta_c$ \\
\hline
Large  & 30.0 & 1.000 & 2.000 & 8.600 & $-0.35$ \\
Medium & 17.6 & 0.873 & 2.255 & 8.769 & $-0.43$ \\
Small  & 11.7 & 0.800 & 2.400 & 7.500 & $-0.45$ \\
\end{tabular}
\end{ruledtabular}
\end{table}

The mean interior density contrast,
\begin{equation}
\label{eq:Delta_bar}
\bar{\Delta}(r) = \frac{3}{r^3}\int_0^r \delta(r')\,r'^2\,dr'\,,
\end{equation}
determines the mean velocity divergence within a sphere of radius
$r$ and enters the monopole formula directly through the streaming
terms.

\subsubsection{Modification to void profiles in $\fR$ gravity}

In Fourier space the $\fR$ void profile is obtained by rescaling
with the growth ratio:
\begin{equation}
\label{eq:delta_fR}
\delta_{f(R)}(k) = \mathcal{R}(k,a)\;\delta_{\rm GR}(k)\,.
\end{equation}
The real-space $\fR$ profile is then obtained by the inverse radial
spherical Bessel transform.  Equivalently, this is the $\ell=0$
component of the spherical Fourier--Bessel transform (or spherical
Hankel transform):
\begin{equation}
\delta_{f(R)}(r) = \frac{1}{2\pi^2}\int_0^\infty
  \mathcal{R}(k,a)\,\delta_{\rm GR}(k)\,\frac{\sin kr}{kr}\,k^2\,dk\,.
\end{equation}
Since $\mathcal{R}(k) > 1$ preferentially at high $k$ (small
scales), the modification is most pronounced in the interior of the
void where the profile has significant small-scale power.

\begin{figure*}[t]
\includegraphics[width=\textwidth]{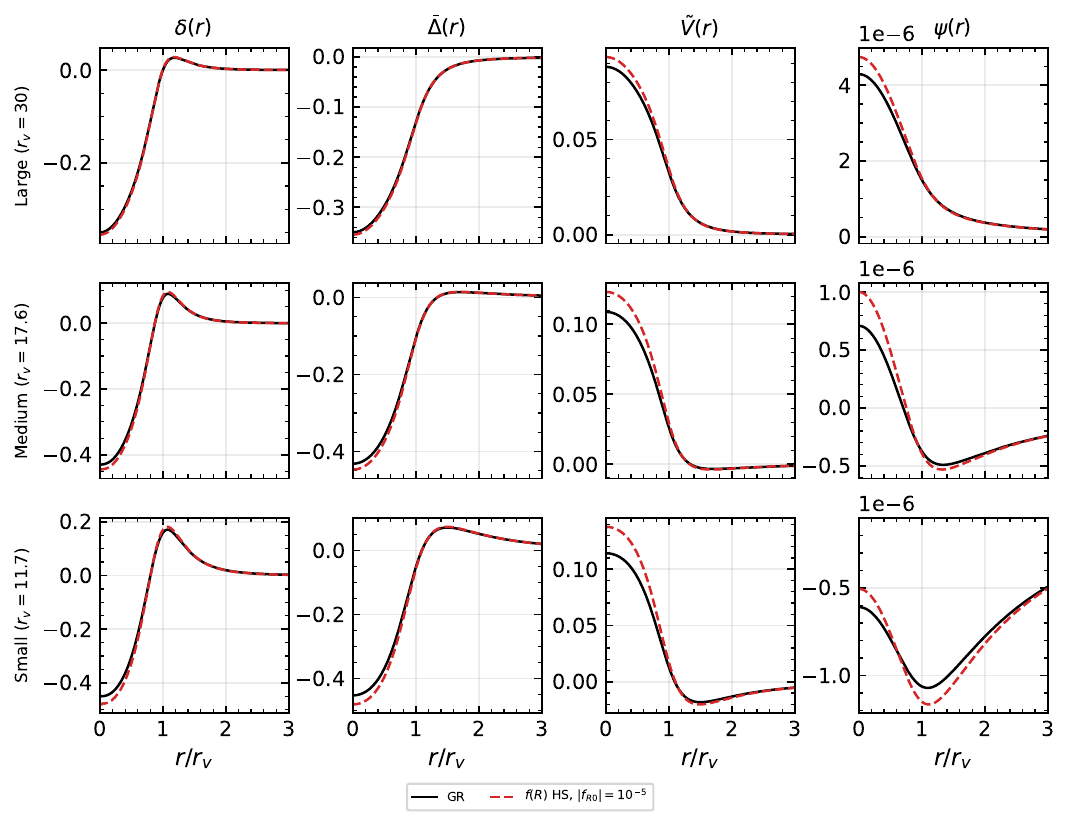}
\caption{Void real-space profiles for $f(R)$ Hu-Sawicki gravity
(red dashed, $\fRz = 10^{-5}$) compared with GR (black solid) at
$z = 0.5$.  Columns from left to right: density contrast $\delta(r)$,
mean enclosed density $\bar{\Delta}(r)$, dimensionless velocity
$\tilde{V}(r)$, and gravitational potential $\psi(r)$.  Rows from top
to bottom: large ($r_v = 30\;\Mpc$), medium ($r_v = 17.6\;\Mpc$), and
small ($r_v = 11.7\;\Mpc$) voids.  The $f(R)$ deviation is most
prominent for small voids (bottom row) due to chameleon unscreening
when $r_v \sim \lambda_C$.  Profile parameters are listed in
Table~\ref{tab:void_params}.}
\label{fig:profiles}
\end{figure*}

\subsection{Velocity Field and Gravitational Potential}
\label{sec:velocity}

\subsubsection{Velocity divergence}

In the linear regime the dimensionless velocity divergence
$\theta(\bm{k},a) \equiv -\nabla\cdot\bm{v}/(aHf)$ satisfies
\begin{equation}
\label{eq:theta_k}
\theta(\bm{k}) = -f(k,a)\,\delta(\bm{k})\,.
\end{equation}
We define the dimensionless radial velocity profile
\begin{equation}
\label{eq:Vtilde}
\tilde{V}(r) = \frac{\bar{\Delta}_\theta(r)}{3}\,,
\end{equation}
where $\bar{\Delta}_\theta(r) = (3/r^3)\int_0^r \theta(r')\,r'^2\,
dr'$ is the mean interior velocity divergence, evaluated with the
same radial spherical Bessel transform from
$\theta(k) = -f(k,a)\,\delta(k)$.

\subsubsection{Gravitational potential}

It is useful to rewrite the modified Poisson equation in terms of the
dimensionless matter density contrast to clarify the notation.  We define
\begin{align}
        \delta\rho_{\rm m}(k,a)
        =
        \bar\rho_{\rm m}(a)\,\delta_{\rm m}(k,a),
        \nonumber
        \\
        \bar\rho_{\rm m}(a)
        =
        {3H_0^2\Omega_{\rm m0}\over 8\pi G}\,a^{-3}.
\end{align}

From now on, for the consistency of notation with RSD dipole analysis, 
we define \(\psi(\bm{k},a)\equiv \Psi(\bm{k},a)\) and \(\delta(\bm{k},a)\equiv\delta_{\rm m}(\bm{k},a)\) 
to denote the gravitational potential and matter density contrast, respectively.  
Starting from Eq.~\eqref{eq:modified_poisson_mu}, in the sub-horizon quasi-static limit, the modified Poisson equation
can be written as
\begin{equation}
\label{eq:poisson_k}
\begin{split}
k^2\,\psi(\bm{k},a)
  &= -A(a)\,\mu_{f(R)}(k,a)\,\delta(\bm{k},a)\,,\\
A(a)&\equiv \frac{3}{2}\frac{\Omega_{\rm m0}H_0^2}{a}\,,
\end{split}
\end{equation}
Equivalently,
one may factor the $f(R)$-modified Fourier potential as
\begin{equation}
\psi_{f(R)}(\bm{k},a)
  = -\frac{A(a)}{k^2}\,\delta_{f(R)}(\bm{k},a)\,\mu_{f(R)}(k,a),
\label{eq:psi_k_yukawa}
\end{equation}
with $\mu_{f(R)}$ given by Eq.~\eqref{eq:mufR}.  Expanding the bracket,
\begin{equation}
\psi_{f(R)}(\bm{k},a)
  = -\frac{A(a)}{k^2}\delta_{f(R)}(\bm{k},a)
    -\frac{A(a)}{3}\,
     \frac{\delta_{f(R)}(\bm{k},a)}{k^2+a^2m_{\rm sc}^2(a)}\, .
\label{eq:psi_k_yukawa_split}
\end{equation}
The first term is the Newtonian/GR-kernel potential sourced by the $f(R)$-modified density contrast, while the second is the scalaron
Yukawa correction.  In real space the GR part can be evaluated using
the Green's function for the spherical Laplacian,
\begin{equation}
\label{eq:psi_r}
\psi_{\rm GR}(r) = -A(a)\left[
  \frac{1}{r}\int_0^r \delta(r')\,r'^2\,dr'
  + \int_r^\infty \delta(r')\,r'\,dr'
\right],
\end{equation}
and the Yukawa correction is
\begin{equation}
\begin{split}
\delta\psi_{\rm Yuk}(\bm{r},a)
  ={}& -\frac{A(a)}{3}
    \int d^3r'\,\delta(\bm{r}',a)\\
   &\times
    \frac{\exp[-a m_{\rm sc}(a)|\bm{r}-\bm{r}'|]}
         {4\pi|\bm{r}-\bm{r}'|}\, .
\end{split}
\label{eq:psi_yukawa_real}
\end{equation}
The transition scale is the scalaron comoving Compton wavenumber
$k_C(a)=a m_{\rm sc}(a)$, or equivalently the comoving reduced Compton length
$\lambda_{C,{\rm com}}(a)=k_C^{-1}(a)$.  In the numerical calculation
we evaluate Eq.~\eqref{eq:poisson_k} in Fourier space and then inverse
transform to real space.

\subsubsection{Yukawa decomposition}

Because $\mu_{f(R)}(k,a)$ decomposes as
$1 + (1/3)\,k^2/(k^2 + a^2m_{\rm sc}^2)$, the $\fR$ potential separates
into a GR piece and the Yukawa correction of
Eq.~\eqref{eq:psi_yukawa_real}:
\begin{equation}
\label{eq:yukawa}
\psi_{f(R)}(r) = \psi_{\rm GR}(r) + \delta\psi_{\rm Yuk}(r)\,.
\end{equation}
The Yukawa piece $\delta\psi_{\rm Yuk}$ is exponentially suppressed on
scales $r \gg \lambda_C$.  For $\fRz = 10^{-5}$ the Compton
wavelength is $\lambda_C \approx 8\;\Mpc$ at $z = 0.5$, so the
correction is significant only in the interior of the void.

The ratio $\psi_{f(R)}(r)/\psi_{\rm GR}(r)$ is therefore
$r$-dependent.  The Fourier-space response has the clean limiting
values $\mu_{f(R)}\to4/3$ for $k\gg k_C$ and
$\mu_{f(R)}\to1$ for $k\ll k_C$, but the real-space potential ratio
is a non-local weighted average over the void density profile and
should not, in general, be identified with the Fourier-space limit
$4/3$.  Its radial dependence reflects the finite Compton range of
the scalaron and is the distinctive Yukawa signature used below.

\begin{figure*}[t]
\includegraphics[width=\textwidth]{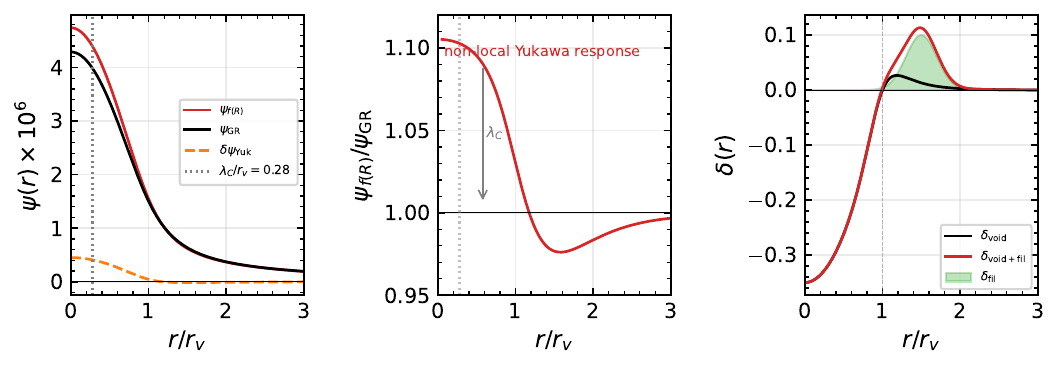}
\caption{Gravitational potential in $f(R)$ gravity for the large void
($r_v = 30\;\Mpc$, $\fRz = 10^{-5}$, $z = 0.5$).
\emph{Left:} Yukawa decomposition---$\psi_{f(R)}$ (red),
$\psi_{\rm GR}$ (black), and the Yukawa correction
$\delta\psi_{\rm Yuk} = \psi_{f(R)} - \psi_{\rm GR}$ (orange dashed),
with the Compton wavelength $\lambda_C/r_v \approx 0.28$ marked
(dotted).
\emph{Center:} ratio $\psi_{f(R)}/\psi_{\rm GR}$, showing the
$r$-dependent Yukawa signature.  This real-space ratio is a non-local
weighted average over the void profile and should not be identified
directly with the Fourier-space unscreened limit $\Geff/G=4/3$.
\emph{Right:} density profile of an isolated void (black) compared
with a void embedded in a filament environment (red), where a Gaussian
ridge at $r_{\rm fil} = 1.5\,r_v$ with width $\sigma_{\rm fil} =
0.2\,r_v$ and amplitude $A_{\rm fil} = 0.1$ models the surrounding
cosmic web (see Appendix~\ref{app:filament}).  The filament ridge
lies at $r \gg \lambda_C$ where screening suppresses the MG
correction, so its primary effect is to modify the compensation
region of the potential.}
\label{fig:potential}
\end{figure*}

\subsection{RSD Multipoles of the Void-Galaxy Cross-Correlation}
\label{sec:rsd}

We follow the formalism of Nan \& Yamamoto~\cite{Nan:2018tce} (see
also~\cite{Kaiser:1987qv,Hamaus:2015gua}) for the void-galaxy
cross-correlation function in redshift space $\xi^s(s,\mu)$, expanding
in Legendre multipoles:
\begin{equation}
\xi^s(s,\mu) = \sum_\ell \xi_\ell(s)\,\mathcal{P}_\ell(\mu)\,,
\end{equation}
where $s$ is the redshift-space separation, $\mu$ the cosine of the
angle to the line of sight, and $\mathcal{P}_\ell$ is the Legendre
polynomial of order $\ell$.  Galaxy bias $b$ enters multiplicatively;
we set $b = 2.0$.

\subsubsection{General structure}

The starting point is the mapping from real-space to redshift-space
coordinates for the void-galaxy cross-correlation.  To distinguish
the full redshift-space correlation $\xi^s(s,\mu)$ from the
underlying radial profile, we write
$\xivg(s) \equiv b\,\delta(s)$ for the biased real-space void-galaxy
correlation profile evaluated at the redshift-space separation~$s$.
The superscript in $\xi^s(s,\mu)$ denotes the full redshift-space
correlation.  Following Ref.~\cite{Nan:2018tce}, the perturbative
mapping $r(s,\mu)$ is expanded around~$s$, so all radial profile
functions ($\delta$, $\xivg$, $\tilde{V}$, $\psi$) are real-space
quantities self-consistently expressed at the observed coordinate~$s$;
the difference $r-s \sim \mathcal{O}(v/\mathcal{H})$ is absorbed into
the streaming correction terms proportional to $\tilde{V}'$,
$\xivg'$, $\xivg''$.
$\tilde{V}(s)$ for the dimensionless radial peculiar velocity, and
$\psi(s)$ for the gravitational potential,
the redshift-space correlation can be expanded in Legendre
multipoles $\xi_\ell(s)$, where all quantities are evaluated at the
redshift-space separation $s$.  The mean enclosed density contrast is
\begin{equation}
\label{eq:Deltabar}
\bar{\Delta}(s) = \frac{3}{s^3}\int_0^s \delta(r')\,r'^2\,dr'\,,
\end{equation}
and primes denote $d/ds$ throughout.

\subsubsection{Monopole}

The monopole ($\ell = 0$) follows from Eq.~(25) of
Ref.~\cite{Nan:2018tce}.  At lowest order in the velocity field the
standard Kaiser-like formula gives
\begin{equation}
\label{eq:xi0_kaiser}
\xi_0^{(0)}(s) = b\,\delta(s) + b\,f\!\left[\bar{\Delta}(s) - \delta(s)\right]
  + \frac{f^2}{3}\left[\bar{\Delta}(s) - \delta(s)\right],
\end{equation}
where $f \equiv d\ln D/d\ln a$ is the growth rate.  The full
expression including the streaming (non-perturbative) corrections
from the coherent velocity field $\tilde{V}$ reads
\begin{align}
\label{eq:xi0_full}
\xi_0(s) &= -1 + (1+\xivg)\Bigl(1 - \tilde{V} + \tilde{V}^2
  - \tfrac{1}{3}\tilde{V}'\,s
  + \tfrac{22}{15}\tilde{V}\tilde{V}'\,s \nonumber\\
  &\quad + \tfrac{1}{5}(\tilde{V}'\,s)^2
  + \tfrac{1}{5}\tilde{V}\tilde{V}''\,s^2\Bigr) \nonumber\\
  &\quad + \tfrac{\xivg'}{15}\,\tilde{V}\,s\,(-5 + 11\tilde{V}
  + 6\tilde{V}'\,s) \nonumber\\
  &\quad + \tfrac{1}{10}\tilde{V}^2\,s^2\,\xivg''\,.
\end{align}
The first line contains the density term and velocity corrections up to
$\mathcal{O}(\tilde{V}^2)$; the second line contributes the streaming
distortion from the gradient $\tilde{V}'$; the third and fourth lines
couple the velocity field to the real-space correlation gradient $\xivg'$
and curvature $\xivg''$.

\subsubsection{Quadrupole}

The quadrupole ($\ell = 2$) arises from the anisotropy between
radial and transverse motions (Eq.~26 of Ref.~\cite{Nan:2018tce}).
Its full expression is
\begin{align}
\label{eq:xi2_full}
\xi_2(s) &= (1+\xivg)\,\tfrac{2}{105}\Bigl(-7\tilde{V}'\,s
  + 29\tilde{V}\tilde{V}'\,s \nonumber\\
  &\quad + 6(\tilde{V}'\,s)^2
  + 6\tilde{V}\tilde{V}''\,s^2\Bigr) \nonumber\\
  &\quad + \tfrac{\xivg'}{105}\,\tilde{V}\,s\,(-14 + 29\tilde{V}
  + 24\tilde{V}'\,s) \nonumber\\
  &\quad + \tfrac{2}{35}\tilde{V}^2\,s^2\,\xivg''\,.
\end{align}
At leading order in $\tilde{V}$ (keeping only $\mathcal{O}(\tilde{V})$
terms), Eq.~\eqref{eq:xi2_full} reduces to
\begin{equation}
\label{eq:xi2_leading}
\begin{split}
\xi_2^{(0)}(s) &= -\frac{2}{15}\,\tilde{V}'\,s\,(1+\xivg)
  - \frac{2}{15}\,\tilde{V}\,s\,\xivg' \\
  &\approx \frac{2\,f}{15}\!\left[\delta(s) - \bar{\Delta}(s)\right]
    + \mathcal{O}(f^2)\,,
\end{split}
\end{equation}
where in the second line we used $\tilde{V}=-f\bar{\Delta}/3$,
$\tilde{V}'s=-f(\delta-\bar\Delta)$, and dropped the
$\mathcal{O}(\xi_{vg})$ bias corrections.
This confirms that the quadrupole is proportional to
$f\,[\delta - \bar{\Delta}]$.  The full streaming expansion
(Eq.~\eqref{eq:xi2_full}) includes higher-order terms that are
numerically important at the $\sim 10\%$ level and are retained
in all our calculations.

\subsubsection{Dipole}

The dipole ($\ell = 1$) uniquely contains a contribution from the
gravitational potential, making it sensitive to the Poisson equation
and hence to $\Geff/G$.  Following Eq.~(27) of
Ref.~\cite{Nan:2018tce}, the dipole splits into velocity and
potential parts:
\begin{equation}
\label{eq:xi1_split}
\xi_1(s) = \xi_1^{\rm vel}(s) + \xi_1^{\psi}(s)\,.
\end{equation}
The velocity part involves the conformal Hubble parameter
$\mathcal{H} = aH$:
\begin{align}
\label{eq:xi1_vel}
\xi_1^{\rm vel}(s) &= (1+\xivg)\Bigl[\tfrac{\mathcal{H}_0 - 3\mathcal{H}}{3}
  \,\tilde{V}^2\,s \nonumber\\
  &\quad + \tfrac{3\mathcal{H}_0 - 11\mathcal{H}}{15}
  \,\tilde{V}\tilde{V}'\,s^2\Bigr] \nonumber\\
  &\quad + \tfrac{3\mathcal{H}_0 - 11\mathcal{H}}{30}
  \,\tilde{V}^2\,s^2\,\xivg'\,,
\end{align}
where $\mathcal{H}_0 = H_0$ is the present-day conformal Hubble
rate.  The potential part is
\begin{equation}
\label{eq:xi1_psi}
\xi_1^{\psi}(s) = \frac{1}{3\mathcal{H}}
  \Bigl[\psi'\,(1+\xivg) + (\psi - \psi_c)\,\xivg'\Bigr]\,,
\end{equation}
where $\psi_c \equiv \psi(r \to 0)$ is the potential at the void
center, ensuring the gauge-invariant
subtraction~\cite{Nan:2018tce,Cai:2016jek}.
The quantity $\psi_c$ is well defined in the stacking procedure:
watershed void finders such as ZOBOV~\cite{Neyrinck:2007gy} identify
void centers as local density minima, and the subsequent radial
stacking in bins of $s = |\bm{s}|$ naturally places $r = 0$ at
the center of each void before averaging.  Because $\psi(r)$ is
smooth near $r = 0$ (with $\psi'(0) = 0$ by spherical symmetry),
the subtraction $\psi - \psi_c$ is numerically stable and does not
introduce additional free parameters.
Throughout we assume that the void velocity field traces the
matter velocity field (velocity bias $b_v = 1$), consistent with
the formulation in Ref.~\cite{Nan:2018tce}.
In $\fR$ gravity the Yukawa-enhanced potential produces a non-local,
$r$-dependent ratio $\psi_{f(R)}(r)/\psi_{\rm GR}(r)$ across the void,
providing a unique diagnostic of the fifth-force range.

\section{Nonlinear Void Evolution}
\label{sec:nonlinear}

The linear treatment of Sec.~\ref{sec:profile} underestimates the
MG signal because the void profile is itself a nonlinear object with
$|\delta| \sim 0.3$--$0.5$ at the void
center~\cite{Bernardeau:1993qu,Sheth:2003py}.  We improve upon this
using shell-by-shell spherical dynamics, following the formalism
developed for spherical collapse and void evolution in the context
of modified gravity~\cite{Clampitt:2012tm,Voivodic:2016plg}.

\subsection{Derivation of the nonlinear shell equation}

Consider a spherical shell initially at Lagrangian radius $q$,
enclosing mass $M(<q) = (4\pi/3)\,\bar{\rho}_i\,(1+\delta_i)\,q^3$.
In the Newtonian limit, the Eulerian radius $R(q,t)$ satisfies
\begin{equation}
\label{eq:Rddot}
\ddot{R} = -\frac{\Geff}{G}\,\frac{G\,M(<q)}{R^2}
  + \frac{\Lambda}{3}\,R\,,
\end{equation}
where the first term includes the modified gravity enhancement via
$\Geff/G$.  Mass conservation relates $R$ to the density contrast:
\begin{equation}
\label{eq:mass_conserve}
(1+\delta)\,R^3 = (1+\delta_i)\,q^3\,\left(\frac{a_i}{a}\right)^3.
\end{equation}
Taking a time derivative of Eq.~\eqref{eq:mass_conserve},
substituting into Eq.~\eqref{eq:Rddot}, and converting from cosmic
time $t$ to redshift $z$ via $d/dt = -(1+z)\,H(z)\,d/dz$, one
obtains (see, e.g., Refs.~\cite{Bernardeau:1993qu,Sheth:2003py,
Voivodic:2016plg} for details of the derivation):
\begin{align}
\label{eq:nl_ode}
\frac{d^2\delta}{dz^2}
  &+ \left(\frac{1}{H}\frac{dH}{dz}
     - \frac{1}{1+z}\right)\frac{d\delta}{dz}  \nonumber\\
  &= \frac{\Geff(k_{\rm eff},a)}{G}\,
     \frac{3\,\Omega_{\rm m0}\,H_0^2}{2\,H^2(z)}\,
     (1+z)\;\delta\,(1+\delta)  \nonumber\\
  &\quad + \frac{4}{3}\,
     \frac{1}{1+\delta}\,
     \left(\frac{d\delta}{dz}\right)^{\!2}.
\end{align}
Equivalently, with $a=1/(1+z)$ and $E(a)\equiv H(a)/H_0$,
Eq.~\eqref{eq:nl_ode} takes the form
\begin{align}
\label{eq:nl_source_a}
\frac{d^2\delta}{da^2}
  &+ \left(\frac{3}{a} + \frac{1}{E}\frac{dE}{da}\right)\frac{d\delta}{da}
= \frac{\Geff(k_{\rm eff},a)}{G}
  \frac{3\,\Omega_{\rm m0}}{2\,a^5\,E^2(a)}\,\delta(1+\delta) \nonumber\\
&\qquad\qquad\quad + \frac{4}{3(1+\delta)}\,\left(\frac{d\delta}{da}\right)^{\!2}\,,
\end{align}
The first line is the standard drag term from the Hubble expansion.
The second line is the gravitational source, where the linear
approximation $\delta$ is replaced by the nonlinear form
$\delta(1+\delta)$, reflecting the fact that the shell encloses
a fixed mass rather than a fixed comoving volume.  The third line is
the velocity self-coupling, arising from the nonlinear relation
between $\dot{R}/R$ and $\dot{\delta}$: the kinetic energy of the
shell feeds back into its deceleration.

The corresponding \emph{linear} equation is recovered by dropping the
two nonlinear corrections:
\begin{equation}
\label{eq:lin_ode}
\frac{d^2\delta}{dz^2}
  + \left(\frac{1}{H}\frac{dH}{dz}
     - \frac{1}{1+z}\right)\frac{d\delta}{dz}
  = \frac{\Geff}{G}\,
     \frac{3\,H_0^2\,\Omega_{\rm m0}}{2\,H^2}\,(1+z)\;\delta\,.
\end{equation}

\subsection{Numerical implementation}

For $\fR$ gravity, $\Geff/G$ is scale-dependent
[Eq.~\eqref{eq:mufR}], so each shell at Lagrangian radius $q$
requires an effective wavenumber to evaluate $\Geff$.  We adopt
\begin{equation}
\label{eq:keff}
k_{\rm eff}(q) = \frac{\pi}{q}\,,
\end{equation}
which is the fundamental mode associated with a perturbation of
spatial extent $\sim q$.

The numerical procedure consists of three steps:
(i)~backscaling the observed void template to $z_{\rm init} = 50$
using the linear growth factor,
(ii)~integrating each of $N_{\rm shell} = 200$ concentric shells
from $z_{\rm init}$ to $z_{\rm target}$ via
Eq.~\eqref{eq:nl_ode}, and
(iii)~recomputing all derived quantities ($\bar{\Delta}$,
$\tilde{V}$, $\psi$) and the RSD multipoles from the nonlinear
profile $\delta_{\rm NL}(r)$.  Full details of the algorithm,
including grid parameters and regularization, are given in
Appendix~\ref{app:algorithm}.

\subsection{Definition of NL amplification}

The key quantity is the NL amplification factor for the MG signal in
a given multipole $\xi_\ell$, defined at the level of the
\emph{observable}:
\begin{equation}
\label{eq:Amp_def}
\mathcal{A}_\ell \equiv
  \frac{\bigl[\Delta\xi_\ell/\xi_\ell^{\rm GR}\bigr]_{\rm NL}}
       {\bigl[\Delta\xi_\ell/\xi_\ell^{\rm GR}\bigr]_{\rm Lin}}\,,
\end{equation}
where $\Delta\xi_\ell \equiv \xi_\ell^{f(R)} - \xi_\ell^{\rm GR}$
and the subscript NL (Lin) indicates that all profiles---density,
velocity, potential---entering the multipole formula are derived from
the nonlinear (linear) evolution.  An amplification $\mathcal{A} > 1$
means that nonlinear evolution enhances the fractional MG signal in
the multipole beyond the linear prediction.

Physically, the NL amplification arises from two nonlinear
corrections in Eq.~\eqref{eq:nl_ode}:
(a)~the gravitational source $\delta(1+\delta)$, which for a void
center with $\delta \approx -0.35$ gives $(1+\delta) \approx 0.65$
--- this reduces the absolute source relative to $\delta$ alone,
but the $\Geff/G$ multiplier acts on the product so that the
\emph{differential} MG-vs-GR signal is enhanced at the profile level;
and
(b)~the velocity self-coupling $(4/3)\,\dot\delta^2/(1+\delta)$,
which grows as $1/(1+\delta)$ for deeper voids and is additionally
enhanced when $\Geff/G > 1$ feeds back through a larger
$|\dot\delta|$.
Decomposing the shell-center MG deviation into these contributions
shows that term~(a) accounts for $\sim 63$--$67\%$ of the NL
\emph{extra} signal (defined as the difference between NL and
linear MG deviations) and term~(b) contributes $\sim 33$--$37\%$.

Importantly, the shell-center NL amplification
$\mathcal{A}_{\rm shell} \approx 0.5$--$0.6$ is actually less than
unity: nonlinear evolution makes voids shallower, which reduces
the MG deviation at the void center.  The observable-level
amplification $\mathcal{A}_\ell > 1$ reported in
Sec.~\ref{sec:nonlinear} therefore arises primarily from the
\emph{multipole formula} nonlinearity --- the functions
$\xi_\ell(s)$ depend on products $\xivg \cdot \tilde{V}$,
$\tilde{V}' \cdot s$, etc., and the derivatives of the NL profile
are steeper near the void wall where the signal peaks.  The
observed amplification is thus a compounding of shell-level effects
(shallower $\delta$ but steeper gradients) with the nonlinear
algebraic structure of the RSD multipole formulas.

\subsection{Order of Magnitude of NL amplification of the RSD multipoles}

\begin{figure*}[t]
\includegraphics[width=\textwidth]{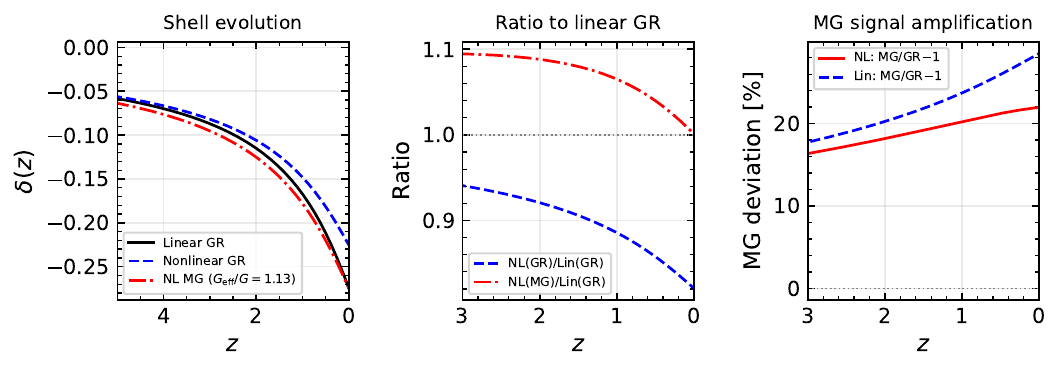}
\caption{Shell-level nonlinear evolution for a void center shell
($\delta_0 \approx -0.35$).
Left: density $\delta(z)$ for linear GR (black solid), nonlinear GR
(blue dashed), and nonlinear MG with $\Geff/G \approx 1.14$ (red
dash-dotted).  Center: ratio to linear GR, showing that at z=0, the nonlinear GR shell is 
about 14\% less underdense than the linear-GR prediction. Right: MG deviation
(percentage difference from GR) in the nonlinear (solid red) vs.\
linear (dashed blue) regimes.  At the shell center, NL evolution
slightly reduces the MG deviation ($\mathcal{A}_{\rm shell} \approx
0.8$ for this depth), consistent with the shallowing effect discussed
in Sec.~\ref{sec:nonlinear}.  The observable amplification
$\mathcal{A}_\ell > 1$ in Fig.~\ref{fig:nl_multipoles} arises from
the nonlinear multipole formulas (products of $\delta$, $\tilde{V}$,
and their gradients), not from enhanced shell-center growth.
The shell equation Eq.~\eqref{eq:nl_ode} is derived from spherical
dynamics~\cite{Bernardeau:1993qu,Sheth:2003py,Voivodic:2016plg}.}
\label{fig:shell_evolution}
\end{figure*}

\begin{figure*}[t]
\includegraphics[width=\textwidth]{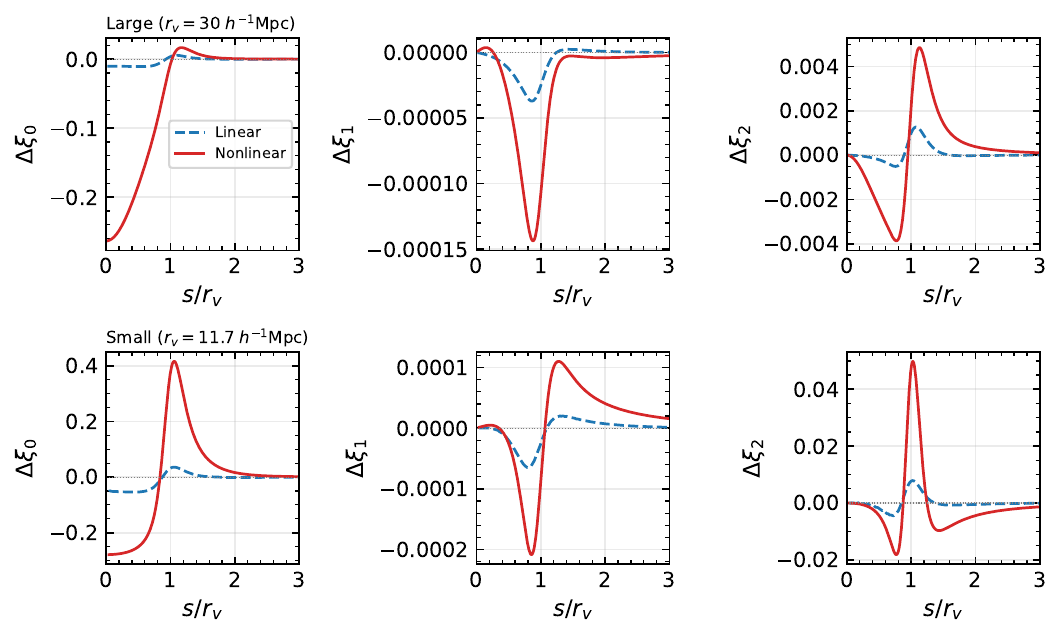}
\caption{NL amplification of RSD multipoles ($|f_{R0}|=10^{-5}$, $z=0.5$).
Blue dashed: linear prediction; red solid: nonlinear shell-by-shell evolution.
\textit{Top row}: large void ($r_v=30\;h^{-1}{\rm Mpc}$) --- monopole $\Delta\xi_0$ ($\mathcal{A}_0\approx3.7$), dipole $\Delta\xi_1$ ($\mathcal{A}_1\approx4.2$), quadrupole $\Delta\xi_2$ ($\mathcal{A}_2\approx4.3$).
\textit{Bottom row}: small void ($r_v=11.7\;h^{-1}{\rm Mpc}$), in the enhanced-response regime --- monopole $\Delta\xi_0$ ($\mathcal{A}_0\approx10$), dipole $\Delta\xi_1$ ($\mathcal{A}_1\approx3.5$), quadrupole $\Delta\xi_2$ ($\mathcal{A}_2\approx5.8$).
The dramatic increase in $\mathcal{A}_0$ from large to small voids reflects the larger fraction of void modes with $\Geff/G > 1$ when $r_v \lesssim \lambda_C$.}

\label{fig:nl_multipoles}
\end{figure*}

Fig.~\ref{fig:shell_evolution} shows the shell-level evolution.
Fig.~\ref{fig:nl_multipoles} shows the NL amplification at the
observable level: the fractional multipole deviation
$\Delta\xi_\ell/\xi_\ell^{\rm GR}$ is plotted for both the linear
and nonlinear treatments.

For the large void class ($r_v = 30\;\Mpc$), we find:
\begin{equation}
\mathcal{A}_0 \approx 3.7\,,\qquad
\mathcal{A}_1 \approx 4.2\,,\qquad
\mathcal{A}_2 \approx 4.3\,.
\end{equation}
All three multipoles are amplified by a comparable factor
$\sim 4$ (with an uncertainty of $\sim \pm 15\%$ from the
$k_{\rm eff}$ prescription; see Appendix~\ref{app:nl_keff}).
The dipole and quadrupole amplifications are
slightly larger than the monopole because they depend more strongly
on the velocity field $\tilde{V}$ and its derivatives, which
compound the NL enhancement of the underlying density profile.

The amplification factor increases for smaller voids: for the
medium (small) void class $\mathcal{A}_0 \approx 5.8$
($10$), reflecting the deeper profiles and stronger
$\Geff/G$ at shorter Compton-scale separations.

Nonlinear evolution thus transforms a linear MG signal in the
monopole into a substantially larger signal that is within reach
of Stage-IV surveys (Sec.~\ref{sec:conclusions}).

\section{Results and Detectability}
\label{sec:results}

\subsection{Fiducial setup and benchmark amplitudes}
\label{sec:fiducial}

We adopt a flat $\lcdm$ background consistent with Planck
2018~\cite{Planck:2018vyg}: $h = 0.6774$, $\Omega_{\rm m0} = 0.3089$,
$\Omega_\Lambda = 0.6911$, and $n_s = 0.9667$.  The MG parameter is
set to $\fRz = 10^{-5}$ unless otherwise stated.  At $z = 0$ this
gives a Compton wavelength $\lambda_C \approx 7.7\;\Mpc$
($\approx 8.3\;\Mpc$ at the fiducial $z = 0.5$).  We have verified
that order-$10\%$ variations in the background cosmological
parameters leave the void-size dependent screening pattern---the
primary observable---unchanged.

Recent cluster-abundance analyses provide strong but model- and
calibration-dependent bounds on the HS amplitude.  For example,
tSZ-selected SPT clusters with DES/HST weak-lensing mass calibration
and Planck 2018 CMB information give
$\log_{10}|f_{R0}|<-5.32$ at the $95\%$ credible level
($|f_{R0}|\lesssim 4.8\times10^{-6}$)~\cite{SPTClusters:2025fR}.
Because this limit relies on cluster mass calibration, selection
functions, and simulation-calibrated halo mass functions, we do not
use it as a hard prior in the present void forecast.  Instead,
$|f_{R0}|=10^{-5}$ is retained as a high-signal benchmark that makes
the void-size dependence and NL amplification transparent, while
$|f_{R0}|=10^{-6}$ represents a conservative amplitude below current
cosmological-scale bounds.  A realistic parameter constraint should
interpolate the template grid and combine void RSD with external
cluster, lensing, and abundance likelihoods rather than imposing a
single external upper limit.

\subsection{Signal shape and void-size dependence}

\begin{figure*}[t]
\includegraphics[width=\textwidth]{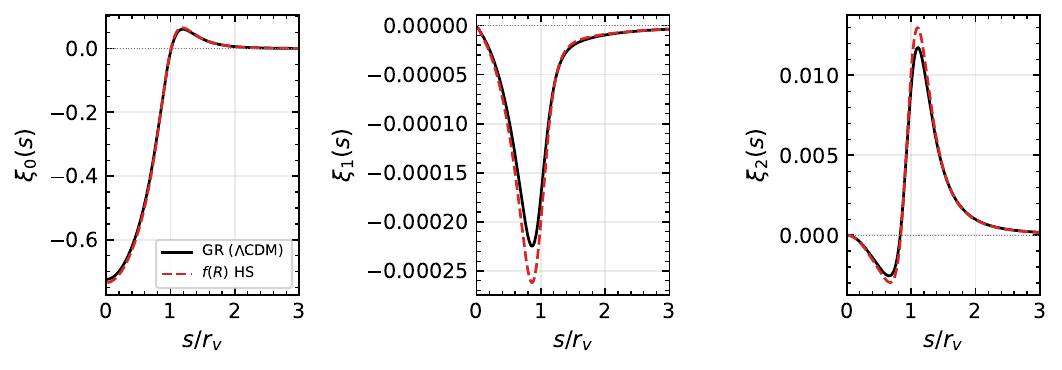}
\caption{RSD multipoles of the void-galaxy cross-correlation at
$z = 0.5$ for the large void ($r_v = 30\;\Mpc$,
$\Delta_c = -0.35$, $\alpha = 2.0$, $\beta = 8.6$).  Black solid:
GR; red dashed: $f(R)$ Hu-Sawicki with $\fRz = 10^{-5}$.  Left:
monopole $\xi_0(s)$; center: dipole $\xi_1(s)$; right: quadrupole
$\xi_2(s)$.}
\label{fig:multipoles}
\end{figure*}

\begin{figure*}[t]
\includegraphics[width=\textwidth]{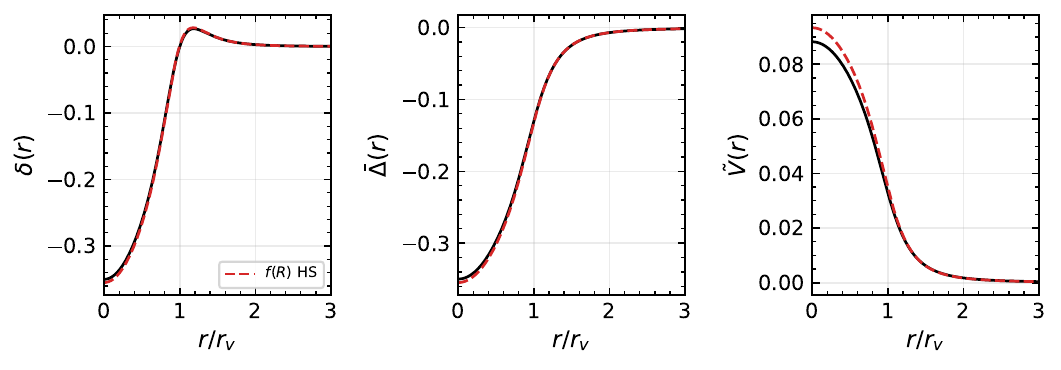}
\caption{Real-space profiles for the large void
($r_v = 30\;\Mpc$, $r_s = 30\;\Mpc$, $\alpha = 2.0$,
$\beta = 8.6$, $\Delta_c = -0.35$) in $f(R)$ Hu-Sawicki gravity
(red dashed, $\fRz = 10^{-5}$) vs.\ GR (black solid) at
$z = 0.5$.  Left: density contrast $\delta(r)$; center: mean
enclosed density $\bar{\Delta}(r)$; right: dimensionless velocity
$\tilde{V}(r)$.  The scale-dependent enhancement from the Yukawa
correction is visible in the velocity profile.}
\label{fig:profiles_fR}
\end{figure*}

\begin{figure*}[t]
\includegraphics[width=\textwidth]{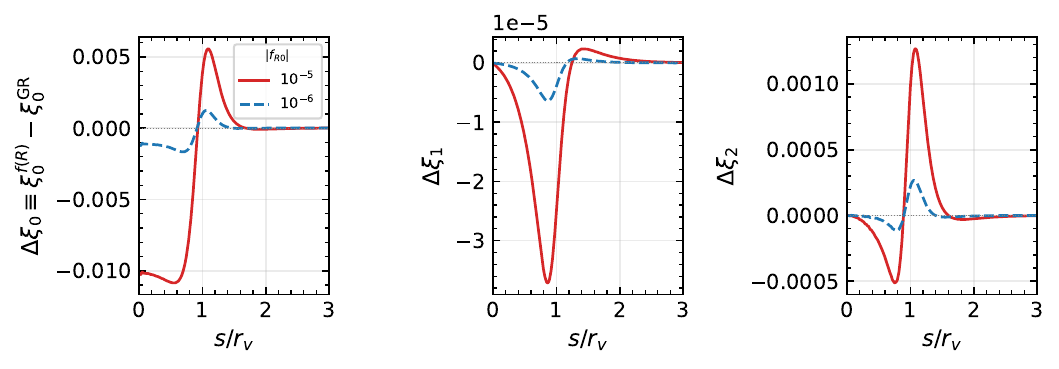}
\caption{Absolute deviation $\Delta\xi_\ell \equiv
\xi_\ell^{f(R)} - \xi_\ell^{\rm GR}$ of the multipoles from GR at
$z = 0.5$ for the large void ($r_v = 30\;\Mpc$).  Red solid:
$\fRz = 10^{-5}$; blue dashed: $\fRz = 10^{-6}$.  Left: monopole
$\Delta\xi_0$; center: dipole $\Delta\xi_1$; right: quadrupole
$\Delta\xi_2$.  The absolute-difference format avoids artificial
divergences at the zero crossings of the GR multipoles.}
\label{fig:deviation_fR}
\end{figure*}

\begin{figure*}[t]
\includegraphics[width=\textwidth]{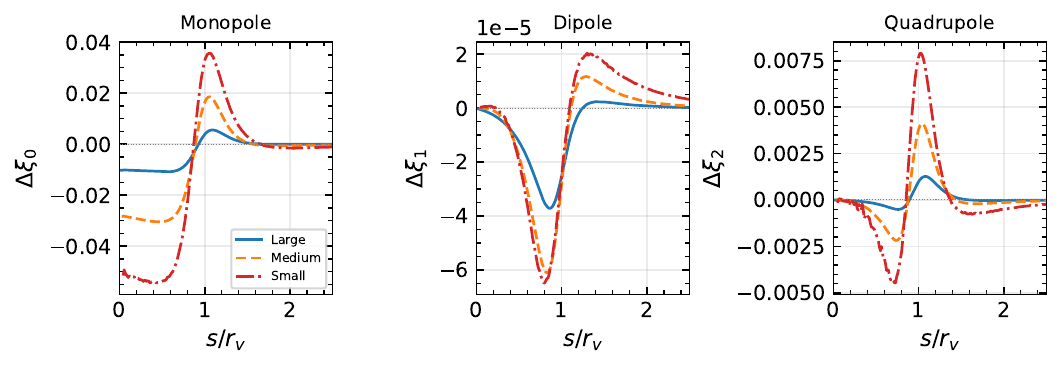}
\caption{Void-size dependence of the absolute multipole deviation
$\Delta\xi_\ell = \xi_\ell^{f(R)} - \xi_\ell^{\rm GR}$ for $f(R)$
Hu-Sawicki gravity ($\fRz = 10^{-5}$, $z = 0.5$).  Left: monopole;
center: dipole; right: quadrupole.  Three void classes are shown:
Large ($r_v = 30\;\Mpc$), Medium ($r_v = 17.6\;\Mpc$), and
Small ($r_v = 11.7\;\Mpc$).
The strong size dependence traces
the scalaron Compton scale: small voids ($r_v \sim \lambda_C
\approx 8\;\Mpc$) contain more sub-Compton modes and show a larger
fifth-force response, while large voids ($r_v \gg \lambda_C$) remain
close to GR in the linear-response approximation.}
\label{fig:size_dep}
\end{figure*}

\begin{figure*}[t]
\includegraphics[width=\textwidth]{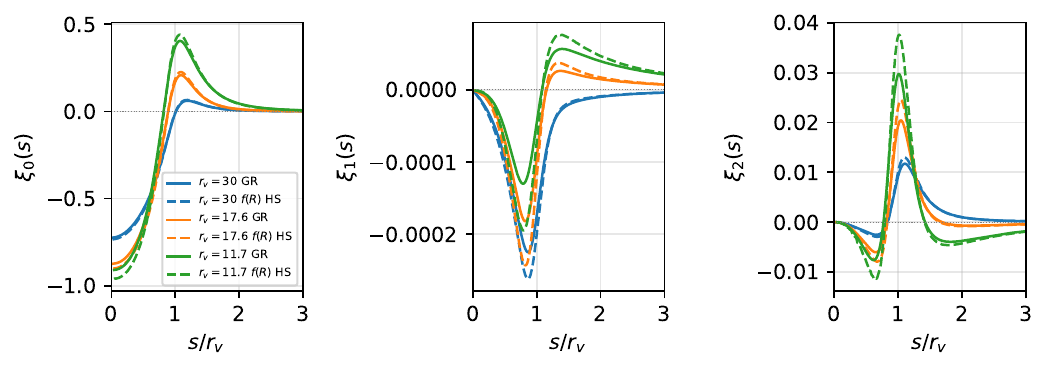}
\caption{RSD multipole curves for all three void size classes at
$z = 0.5$ with $\fRz = 10^{-5}$.  Solid lines: GR; dashed lines:
$f(R)$ Hu-Sawicki.  Left: monopole $\xi_0(s)$; center: dipole
$\xi_1(s)$; right: quadrupole $\xi_2(s)$.  The separation between
GR and $f(R)$ curves grows dramatically for smaller voids, directly
reflecting the void-size dependence in Table~\ref{tab:results}.}
\label{fig:multipoles_3sizes}
\end{figure*}

Table~\ref{tab:results} summarizes the peak \emph{linear} fractional
deviation of the monopole $\xi_0(s)$ from $\lcdm$ for the three void
size classes, computed from scale-dependent linear profiles (no NL
shell evolution; see also Eq.~\eqref{eq:delta_fR} and
Sec.~\ref{sec:nonlinear}).  The deviation increases dramatically for
smaller voids: from $+2.8\%$ at $r_v = 30\;\Mpc$ to $+29.7\%$ at
$r_v = 11.7\;\Mpc$.

\begin{table}[t]
\caption{Peak \emph{linear} fractional deviation of the monopole,
$\xi_0^{f(R)}/\xi_0^{\rm GR} - 1$, evaluated at its maximum
in the void interior region ($\xi_0^{\rm GR} < 0$),
for $\fRz = 10^{-5}$ at $z = 0.5$, computed using scale-dependent
linear profiles (no NL shell evolution).
The NL monopole amplification factor $\mathcal{A}_0$
(as defined in Eq.~\eqref{eq:Amp_def}) is also given;
the NL peak deviation is $\mathcal{A}_0$ times larger.
Multipole curves are shown in Fig.~\ref{fig:multipoles_3sizes}.}
\label{tab:results}
\begin{ruledtabular}
\begin{tabular}{lccc}
Void class & $r_v\;[\Mpc]$ & Deviation (\%) & $\mathcal{A}_0$ \\
\hline
Large  & 30.0  & $+2.8$  & $3.7$ \\
Medium & 17.6  & $+8.0$  & $5.8$ \\
Small  & 11.7  & $+29.7$ & $10.2$ \\
\end{tabular}
\end{ruledtabular}
\end{table}

The physical origin of this void-size dependence is the
Compton-scale scalaron response associated with chameleon screening.
Large voids have $r_v \gg \lambda_C \approx 8\;\Mpc$, so the density
profile is dominated by Fourier modes with $k \lesssim a m_{\rm sc}$ that are
close to the GR regime.  Small voids have $r_v \sim \lambda_C$, and a
larger fraction of their Fourier content lies at $k \gtrsim a m_{\rm sc}$
where $\Geff/G$ is significantly enhanced.  In the present
linear-response treatment, $\lambda_C$ therefore sets the transition
between effectively screened and unscreened void-size bins.  A full
chameleon calculation may shift the precise transition scale because
the scalaron mass depends on the local density and environment.

For the more stringent $\fRz = 10^{-6}$, the Compton wavelength
shrinks to $\lambda_C \approx 2.6\;\Mpc$ and the monopole deviation
drops to $+0.3\%$ (large voids) and $+3.1\%$ (small voids).
The growth ratio $\mathcal{R}(k,a)$ approaches unity across most of
the $k$-range, confirming that $\fRz = 10^{-6}$ is near the edge of
detectability with void-RSD (see Sec.~\ref{sec:detectability}).

\subsubsection{Quadrupole and nonlinear amplification}

The quadrupole $\xi_2(s)$ depends quadratically on the growth rate
($\propto f^2$) and on the velocity field $\tilde{V}$ and its
derivatives.  For the large void class the nonlinear amplification
of the quadrupole deviation is $\mathcal{A}_2 \approx 4.3$.
The monopole amplification is $\mathcal{A}_0 \approx 3.7$.
For smaller voids, $\mathcal{A}_0$ rises to $\sim 5.8$--$10$ (see
Appendix~\ref{app:nl_keff}, Table~\ref{tab:nl_all}).

\begin{figure}[tb]
\includegraphics[width=\columnwidth]{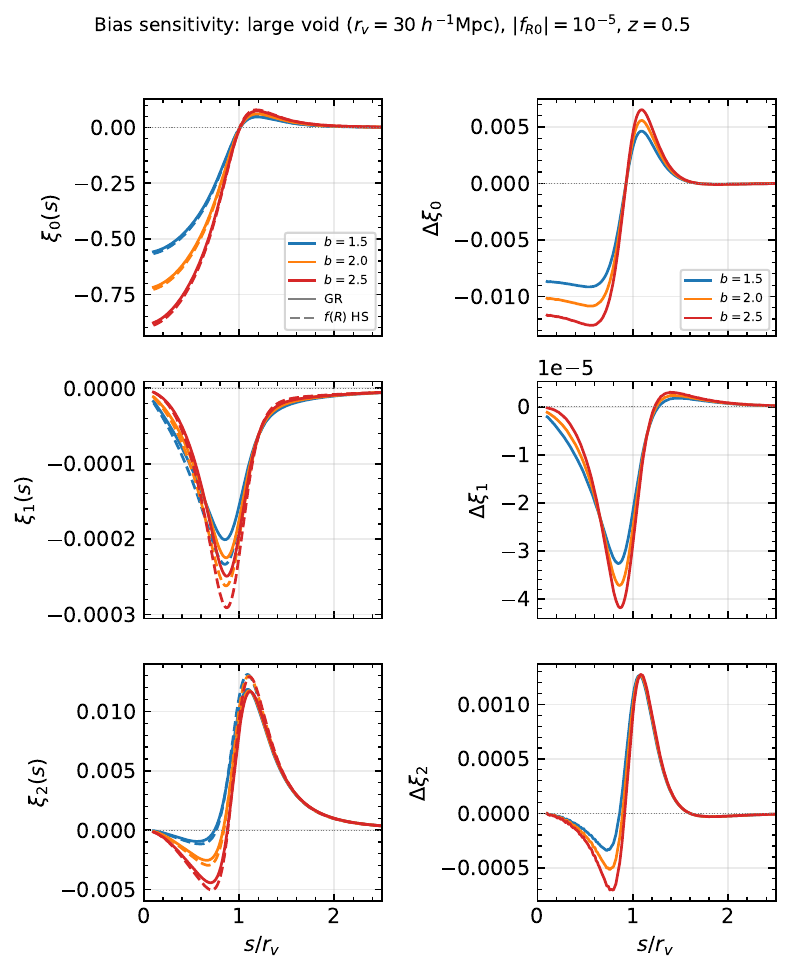}
\caption{Galaxy bias sensitivity of the RSD multipole MG signal
($r_v = 30\;\Mpc$, $\fRz = 10^{-5}$, $z = 0.5$).  Left column:
absolute multipoles (GR solid, $\fR$ HS dashed); right column:
absolute MG deviation $\Delta\xi_\ell = \xi_\ell^{f(R)} -
\xi_\ell^{\rm GR}$.  Rows: monopole $\xi_0$ (top), dipole $\xi_1$
(middle), quadrupole $\xi_2$ (bottom).  Colors: $b=1.5$ (blue),
$b=2.0$ (orange), $b=2.5$ (red).  The quadrupole and dipole MG
deviations are nearly insensitive to $b$; the monopole shows
$\lesssim 1.5\%$ variation across the bias range.}
\label{fig:bias_sens}
\end{figure}

Importantly, the quadrupole fractional MG deviation is nearly
independent of the galaxy bias parameter $b$.
Scale-dependent and radially varying bias effects inside voids have
been studied in N-body simulations~\cite{Pollina:2017,Hamaus:2014afa},
with the conclusion that a constant linear bias is a good approximation
for the void-galaxy cross-correlation at the profile scales
$r \gtrsim 5\;\Mpc$ considered here.  To assess the residual
sensitivity, Fig.~\ref{fig:bias_sens} shows the monopole and
quadrupole deviation for $b = 1.5$, $2.0$, and $2.5$.  Varying $b$
across this range changes the quadrupole absolute MG deviation
$\Delta\xi_2$ by $\lesssim 0.5\%$ of its peak value, whereas the
monopole $\Delta\xi_0$ shifts by $\lesssim 1.5\%$ of its peak.
This $b$-insensitivity of $\xi_2$ arises because the quadrupole is
dominated by velocity terms ($\propto f^2$) that do not involve $b$,
making it a robust MG diagnostic with minimal galaxy-bias systematics.

\subsubsection{Dipole and the Yukawa potential}

The dipole $\xi_1(s)$ contains the gravitational potential term
$\xi_1^{\psi}$.  In $\fR$ gravity the ratio
$\psi_{f(R)}(r)/\psi_{\rm GR}(r)$ is $r$-dependent due to the
finite-range Yukawa correction.  Unlike the Fourier-space response,
whose unscreened limit is $\mu_{f(R)}\to4/3$, this real-space ratio
is a weighted convolution over the void density profile and is not a
pointwise $4/3$ rescaling.  This $r$-dependent enhancement is a direct
diagnostic of chameleon screening.  In contrast, in theories without a
screening mechanism the potential ratio is approximately constant
across the void.  The dipole thus provides a way to discriminate
between screened and unscreened classes of modified gravity theories.

\subsection{Redshift dependence}
\label{sec:zdep}

\begin{figure*}[t]
\centering
\includegraphics[width=\textwidth]{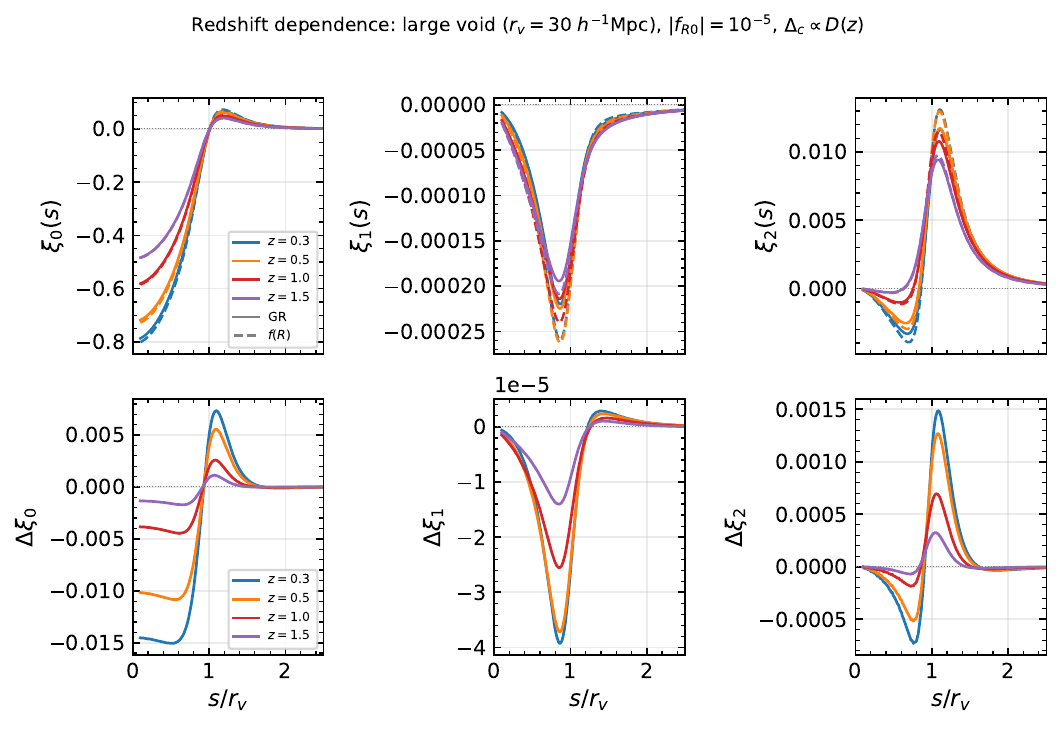}
\caption{Redshift dependence of the RSD multipoles for the large
void ($r_v = 30\;\Mpc$, $\fRz = 10^{-5}$).
Top: GR (solid) vs.\ $\fR$ (dashed) multipoles at four redshifts.
Bottom: absolute MG deviation $\Delta\xi_\ell = \xi_\ell^{f(R)} -
\xi_\ell^{\rm GR}$.  The Compton wavelength at each redshift is
$\lambda_C(z{=}0.3) = 8.4$, $\lambda_C(z{=}0.5) = 8.3$,
$\lambda_C(z{=}1.0) = 6.9$, $\lambda_C(z{=}1.5) = 4.9$~$\Mpc$.
The signal decreases at higher~$z$ as the
growth factor diminishes and the Compton wavelength shrinks.
Note that the void profile template is held fixed across
redshifts (Hamaus et al.\ $z \approx 0.5$ parameters).}
\label{fig:zdep}
\end{figure*}

Fig.~\ref{fig:zdep} shows the RSD multipoles and their MG deviations
at $z = 0.3$, $0.5$, $1.0$, and $1.5$ for the large void class.
The MG signal decreases monotonically with increasing redshift:
the peak monopole deviation drops by a factor of $\sim 6.5$ from
$z = 0.3$ to $z = 1.5$.
Two competing effects govern this evolution.
On one hand, the Compton wavelength $\lambda_C$ is relatively flat
at low $z$ ($\lambda_C \approx 8.4\;\Mpc$ at $z = 0.3$;
$8.3\;\Mpc$ at $z = 0.5$) but decreases at high~$z$
($6.9\;\Mpc$ at $z = 1.0$; $4.9\;\Mpc$ at $z = 1.5$)
as the scalaron mass grows with the background curvature.
On the other hand, the linear growth factor $D(z)$ is smaller at
higher~$z$, reducing the amplitude of the density perturbation
and hence the absolute MG deviation.
A third effect (not included in Fig.~\ref{fig:zdep}, which uses
the linear treatment only) is the reduction of nonlinear amplification
at high~$z$: the smaller density perturbations $\delta(z)$ weaken
the nonlinear source $\delta(1+\delta)$ in the shell ODE, so
$\mathcal{A}_0$ is expected to decrease from its $z = 0.5$
value of $\sim 4$.  The NL S/N estimates in Table~\ref{tab:sn_nl}
are calibrated at $z=0.5$ only; extrapolating them to higher
redshift would require recomputing the shell evolution at each~$z$.
At the redshifts probed by DESI ($z \sim 0.5$--$1.0$) and
Euclid ($z \sim 1.0$), the signal remains well above the
detection threshold (Sec.~\ref{sec:detectability}).
Multi-redshift void-RSD measurements can map $\lambda_C(z)$
and provide an independent constraint on the scalaron mass
function $m_{\rm sc}^2(a)$.
We note that the void profile template parameters used here are
calibrated at $z \approx 0.5$~\cite{Hamaus:2014afa}; at $z > 1$
voids are typically shallower ($|\Delta_c|$ smaller) with
different shape parameters, which could modify the absolute
deviation at the $20$--$30\%$ level.  The trends in Fig.~\ref{fig:zdep}
should therefore be interpreted as isolating the effects of
$\lambda_C(z)$ and $D(z)$ at fixed profile shape, rather than
as full predictions at each redshift.  In particular, the
profile shape parameters ($\alpha$, $\beta$, $r_s/r_v$) may also
evolve with redshift as void populations mature, which would
further modulate the $\lambda_C$-to-multipole mapping at high~$z$.

\subsection{Detectability forecasts}
\label{sec:detectability}

We assess the prospects for detecting both the RSD multipole signals
themselves and the $\fR$-vs-GR deviation with current and upcoming
surveys.

We adopt the survey parameters and phenomenological covariance model
described in Appendix~\ref{app:covariance}
(Table~\ref{tab:surveys}).  The cumulative S/N for the $\fR$-vs-GR
deviation is
\begin{equation}
\mathrm{S/N} =
  \left[\Delta{\bf d}^{\,T} C_{\rm syn}^{-1}\Delta{\bf d}\right]^{1/2},
\label{eq:sn_cum}
\end{equation}
where ${\bf d}_a=\xi_\ell(s_i)$ with $a=(\ell,i)$, and the fitted
radial range is $5\,h^{-1}{\rm Mpc}\leq s\leq 2\,r_v$; throughout this paper
all quoted S/N values are the signal-to-noise amplitude, i.e., the
square root of the quadrature sum.

\subsubsection{MG discrimination: $\fR$ vs.\ GR}
\label{sec:detect_mg}

We adopt $\mathrm{S/N} \geq 3$ (the $3\sigma$ criterion) as the
threshold for a confident detection of the MG signal.  Values below
this threshold do not indicate the model is ruled out; rather, a
non-detection at measured $\mathrm{S/N} = x < 3$ places an upper
bound on the MG parameter after interpolating the $\fRz=10^{-5}$
and $10^{-6}$ templates.  A simple power-law rescaling is only
indicative: the Compton wavelength scales as $\lambda_C\propto
\fRz^{1/2}$~\cite{Hu:2007nk}, whereas the fixed-$k$ linear response
and the RSD multipoles do not obey a single universal power law.
Thus a non-detection still provides a competitive \emph{constraint},
but the quoted limits should be obtained from the template grid
rather than from a single analytic scaling.

\textit{Nonlinear S/N estimates.}---The linear template S/N values
are computed from the \emph{linear} MG deviation $\Delta\xi_\ell^{\rm lin}$.
The NL amplification factors $\mathcal{A}_\ell$ in
Appendix~\ref{app:nl_keff} are defined for the fractional multipole
deviation, not for the full survey likelihood.  If one keeps the same
synthetic covariance model and rescales only the signal amplitude, an
indicative estimate is
$\mathrm{S/N}_{\rm NL} \simeq \mathcal{A}_0\,\mathrm{S/N}_{\rm lin}$
for the monopole.  Table~\ref{tab:sn_nl} gives this upper-envelope
estimate.  A definitive nonlinear forecast must recompute the
covariance, void abundance, and profile scatter from nonlinear
$\fR$ mock catalogs; the values below should not be interpreted as a
replacement for that analysis.

The linear synthetic-covariance calculation gives a robust hierarchy:
the MG discrimination is monopole dominated, small and medium voids
carry the strongest response, and the quadrupole and dipole provide
complementary diagnostic information in this compressed covariance
model.  For example, with the baseline survey counts of
Table~\ref{tab:surveys}, the linear monopole S/N at
$|f_{R0}|=10^{-5}$ is $\sim 4.6$--$6.6$ for DESI~Y5,
$\sim 9.2$--$13.1$ for the Roman reference HLSS forecast, and
$\sim 7.3$--$10.4$ for the DESI+Euclid+PFS combined sample across the
three void-size bins.  At $|f_{R0}|=10^{-6}$ the corresponding
small-void values are $\sim 1.5$ (DESI~Y5), $\sim 3.0$ (Roman reference HLSS),
and $\sim 2.4$ (DESI+Euclid+PFS combined).
Because the Roman reference HLSS forecast probes $z = 1$--$3$, where the Compton
wavelength is shorter, its contribution to the combined MG
discrimination S/N is modest: a quadrature combination of all
surveys (including Roman) increases the linear monopole by only
$\sim 5$--$10\%$ beyond the DESI+Euclid+PFS combined sample.
Roman nonetheless provides an independent high-redshift lever arm
that can map the $\lambda_C(z)$ evolution.
These values should be read as fixed-template detectability estimates,
not as a marginalized survey likelihood.

\begin{table}[tb]
\caption{NL-corrected MG-discrimination S/N for the monopole
$\xi_0$ ($z = 0.5$), estimated as
$\mathrm{S/N}_{\rm NL} \simeq \mathcal{A}_0 \times \mathrm{S/N}_{\rm lin}$
using the amplification factors $\mathcal{A}_0 \approx 3.7$ (large),
$5.8$ (medium), $10$ (small) from Appendix~\ref{app:nl_keff}.
The underlying linear S/N values use the synthetic non-diagonal
covariance of Appendix~\ref{app:covariance}; no additional
``diagonal-to-full'' degradation factor is applied.
The Roman row corresponds to the 2\,000\,deg$^2$ reference HLSS forecast,
while the Combined row denotes the DESI+Euclid+PFS bookkeeping aggregate.
A full nonlinear analysis requires mock catalog covariance calibration,
so the quoted values should be read as illustrative template-level
compressed estimates rather than forecast-level marginalized constraints.}
\label{tab:sn_nl}
\begin{ruledtabular}
\footnotesize
\begin{tabular}{l cc cc cc}
  & \multicolumn{2}{c}{Small}
  & \multicolumn{2}{c}{Medium}
  & \multicolumn{2}{c}{Large} \\
Survey & $10^{-5}$ & $10^{-6}$ & $10^{-5}$ & $10^{-6}$
       & $10^{-5}$ & $10^{-6}$ \\
\hline
BOSS     & 13 &  4 & 10 &  3 &  4 & 1 \\
DESI Y5  & 46 & 15 & 38 & 10 & 14 & 3 \\
Euclid   & 38 & 12 & 31 &  9 & 11 & 2 \\
Subaru PFS & 23 & 7 & 19 & 5 & 7 & 1 \\
Roman    & 92 & 30 & 76 & 21 & 27 & 5 \\
Combined & 73 & 24 & 60 & 17 & 21 & 4 \\
\end{tabular}
\end{ruledtabular}
\end{table}

\subsubsection{Direct multipole detection and compressed estimators}

It is useful to distinguish the detection of a multipole itself from
the detection of the $\fR$-vs-GR difference in that multipole.
Table~\ref{tab:sn_direct} shows the direct S/N of the GR multipoles
for the large-void template.  The monopole is overwhelmingly measured,
and the quadrupole should be directly detectable in DESI Y5, Euclid,
and the DESI+Euclid+PFS combined sample.  The uncompressed dipole is
smaller in this baseline estimator, with S/N below unity, but this
should be interpreted as a statement about the simple estimator and
baseline void counts rather than as a fundamental no-go result.
Since the covariance scales approximately as $1/N_v$, raw S/N values
increase roughly as $N_v^{1/2}$ when the same void selection can be
extended to larger effective catalogs.

\begin{table}[tb]
\caption{Direct-detection S/N for the multipoles themselves in the
large-void GR template at $z=0.5$, using the same synthetic covariance
model and radial range as Eq.~\eqref{eq:sn_cum}.  These are not
$\fR$-vs-GR discrimination significances.  The corresponding
$|f_{R0}|=10^{-5}$ values differ only slightly at the precision shown.
The Roman row corresponds to the 2\,000\,deg$^2$ reference HLSS forecast,
while the Combined row denotes the DESI+Euclid+PFS bookkeeping aggregate.}
\label{tab:sn_direct}
\begin{ruledtabular}
\begin{tabular}{lccc}
Survey & $\xi_0$ & $\xi_1$ & $\xi_2$ \\
\hline
BOSS          &  35 & $<0.1$ & 0.7 \\
DESI Y5       & 127 & $<0.1$ & 2.6 \\
Euclid        & 103 & $<0.1$ & 2.1 \\
Subaru PFS    &  62 & $<0.1$ & 1.3 \\
Roman         & 251 & 0.1    & 5.1 \\
Combined      & 200 & 0.1    & 4.1 \\
\end{tabular}
\end{ruledtabular}
\end{table}

For the dipole, the MG discrimination S/N is additionally suppressed
because the gravitational-potential contribution carries the factor
$1/(3{\cal H})$ and partly cancels between
$\psi'$ and $(\psi-\psi_c)\xivg'$.
Even the direct detection of the GR dipole itself is challenging:
the absolute $\xi_1$ S/N reaches only $\sim 0.1$ for the Roman reference HLSS
($N_v > 8\times 10^4$, $z \sim 2$) and $\sim 0.1$ for the
quadrature combination of all surveys including Roman---still well
below $3\sigma$.  In this uncompressed estimator a significant
dipole measurement would require orders of magnitude more effective
voids, unless the radial template and external fields are exploited.
The realistic path is therefore
a compressed or external-template estimator:
matched filtering of the predicted radial dipole shape, joint
fitting of $\xi_1$ with void-CMB-lensing or ISW templates that
isolate the potential field, and the use of the full Hamaus-style
covariance rather than independent radial bins.  This strategy is
analogous to the CMB-lensing/void matched-filter measurements used
for superstructure profiles and to the potential-isolating proposal
of Ref.~\cite{Cai:2016jek}.  We therefore treat the dipole as a
high-value consistency test of the Yukawa potential and a promising
target for compressed estimators, while the monopole carries the main
stand-alone MG detection power in the present semi-analytical forecast.

We caution that these are order-of-magnitude estimates.  Additional
sources of uncertainty---sample/cosmic variance of the void
population, systematic dependence on void-finder algorithm and radius
definition~\cite{Nadathur:2019mct}, off-diagonal radial-bin
covariance, and void overlap---could reduce the effective S/N by a
factor of $\sim 2$--$3$ (see Appendix~\ref{app:covariance} for
details).

\subsubsection{Optimistic prospects}
\label{sec:optimistic}

The conservative single-size, single-redshift estimates above represent
a \emph{lower bound} on the achievable sensitivity.  Several
strategies can substantially improve the MG discrimination power:

\emph{(i)~Multi-size stacking.}---Within the same synthetic-covariance
bookkeeping used above, one can treat the three void-size bins
(small, medium, large) as approximately independent and add their
S/N values in quadrature:
$\mathrm{S/N}_{\rm tot} = [\sum_i (\mathrm{S/N}_i)^2]^{1/2}$.
Combining the three bins at $|f_{R0}|=10^{-5}$ yields
$\mathrm{S/N}_{\rm tot}\sim 1.5$--$2\times$ the single-bin value,
with the enhancement driven by the small-void bin where the
screening transition maximizes the signal.

\emph{(ii)~Multi-redshift binning.}---Future surveys (DESI~Y5, Euclid)
span $0.2 \lesssim z \lesssim 1.6$ and can be sliced into
$N_z \sim 4$--$6$ redshift bins.  If those bins are treated as
approximately independent at the same template level, the gain is $\sim\!\sqrt{N_z}$, giving a factor of
$2$--$2.5$ beyond a single effective-$z$ analysis, while
simultaneously mapping the evolution of the Compton wavelength
$\lambda_C(z)$.

\emph{(iii)~Template matched filtering.}---The S/N formula
Eq.~\eqref{eq:sn_cum} uses a fixed compressed covariance model.
An optimal matched filter that down-weights noisy bins (near zero
crossings, at large $s$ where shot noise dominates) and up-weights
the peak-signal region can improve the effective S/N by a factor
$\sim 1.3$--$1.5$ over uniform binning, as demonstrated for
void-lensing stacks in Ref.~\cite{Baker:2019gxo}.

\emph{(iv)~Cross-probe synergies.}---Joint fitting of void-galaxy
RSD with void lensing~\cite{Baker:2019gxo} or ISW
stacking~\cite{Cai:2016jek} breaks the $\Geff$--$b$--$f$ degeneracy
and effectively replaces the external-$b$ prior, tightening
constraints by an additional factor that depends on the lensing
signal-to-noise.

We caution that these gains are \emph{not} multiplicatively
independent: multi-size and multi-$z$ bins share cosmic variance
within overlapping survey volumes, and the matched-filter
improvement depends on the assumed noise model.  A realistic
estimate of the combined improvement is $\sim 2$--$3\times$
rather than the na\"ive product $\sqrt{3}\times\sqrt{5}\times 1.4
\approx 6$.  Even this modest combined factor would push the
$3\sigma$ detection threshold at $|f_{R0}|=10^{-6}$ from marginal
(single-bin DESI~Y5 linear S/N~$\sim 1.5$--$1.8$) toward the
detectable regime, motivating a full
simulation-calibrated likelihood analysis as the next step.

\section{Discussion and Conclusions}
\label{sec:conclusions}

We have developed a semi-analytical theory framework for computing
the monopole, dipole, and quadrupole of the void-galaxy
cross-correlation function in redshift space for the Hu-Sawicki
$\fR$ model with $\fRz = 10^{-5}$.  Our main findings are as
follows.

\textit{Void-size dependence as a hallmark of the scalaron Compton scale.}---
The MG deviation in the monopole ranges from $+2.8\%$ for large voids
($r_v = 30\;\Mpc$) to $+29.7\%$ for small voids ($r_v = 11.7\;
\Mpc$).  This strong size dependence is a direct consequence of the
Compton wavelength $\lambda_C \approx 8\;\Mpc$ (at $z = 0.5$) of the
scalaron: voids with Fourier support at $k \gtrsim a m_{\rm sc}$ feel the
enhanced force, while larger voids remain close to the GR response.
Measuring the void-size dependence of the RSD multipoles would
therefore test the Compton-scale transition expected in chameleon
$\fR$ gravity~\cite{Falck:2017rvl,Tamosiunas:2022csc}.  The exact
environmental screening threshold should be calibrated with
simulations that solve the nonlinear scalaron field.

\textit{Nonlinear amplification.}---Nonlinear spherical shell
evolution amplifies the MG signal at the observable (multipole) level
by a factor of $\mathcal{A}_0 \approx 3.7$ for the monopole and
$\mathcal{A}_2 \approx 4.3$ for the quadrupole (large void).
This amplification arises because the nonlinear source and velocity
corrections both involve $\Geff/G$, and it increases further for
smaller voids ($\mathcal{A}_0 \sim 6$--$10$).

\textit{Yukawa potential as a discriminant.}---The $r$-dependent ratio
$\psi_{f(R)}/\psi_{\rm GR}$ encodes the Yukawa profile of the
scalaron-mediated fifth force; in theories without screening the
ratio is nearly flat.  The dipole of the void-galaxy
cross-correlation, which depends directly on $\psi$, provides a
clean observable for this diagnostic~\cite{Nan:2018tce}.
We note, however, that the absolute dipole signal is below the
nominal detection threshold with current survey designs
($\mathrm{S/N}<1$ for MG discrimination in the present
synthetic covariance model), so
exploiting this diagnostic will require either substantially
larger void catalogs or cross-correlation with CMB
lensing/ISW maps~\cite{Cai:2016jek}.

\textit{Generality beyond the Hu-Sawicki model.}---Although we focus
on the HS $n = 1$ model as a well-motivated benchmark, the
semi-analytical framework is not restricted to it.  The growth
equation Eq.~\eqref{eq:growth_ode}, the modified Poisson
equation Eq.~\eqref{eq:poisson_k}, and the shell
equation Eq.~\eqref{eq:nl_ode} require only a specification of
$\Geff(k,a)/G$.  Any metric $\fR$ theory that admits a quasi-static
limit yields a $\Geff$ of the form
$1 + F(k,a)/(3[1 + F(k,a)])$~\cite{DeFelice:2010aj}, where
$F = k^2/(a^2 m_{\rm sc}^2)$ for the HS model.  The Starobinsky
model~\cite{Starobinsky:2007hu}, designer $f(R)$
models~\cite{Song:2006ej}, and the general chameleon
class~\cite{Brax:2012gr} all fit this structure.  The void-size
dependent screening signature is generic to any theory with a
scale-dependent $\Geff$; what changes between models is the
functional form of $m_{\rm sc}^2(a)$ and hence the Compton wavelength
$\lambda_C$.  Our results can therefore be reinterpreted for
other $\fR$ models by simply substituting the appropriate
$m_{\rm sc}^2(a)$ into Eq.~\eqref{eq:mufR}.

\textit{Connection to dark energy clustering.}---Metric $f(R)$ gravity is dynamically equivalent to a
Brans--Dicke-type scalar--tensor theory with
$\omega_{\rm BD}=0$ and a nontrivial scalaron potential.
The scalaron mediates a finite-range fifth force whose
strength changes across the Compton scale, producing a
characteristic void-size dependence in the void-RSD signal.
In the equivalent effective-fluid description, the same
scale-dependent response corresponds to spatial perturbations
of the effective dark-energy component.  Thus, the
void-size-dependent transition in void-RSD can be interpreted
not only as a test of modified gravity, but also as a probe of
effective dark-energy clustering on sub-horizon scales.
\cite{DeFelice:2010aj,Sotiriou:2008rp,Nan:2019,Nan:2022,Takada:2006xs}.

\textit{Comparison with other void probes.}---Our analytical results
are complementary to void abundance
studies~\cite{Voivodic:2016plg,Perico:2019obq,Sahlen:2018dks,Contarini:2023qqx}
and void lensing
analyses~\cite{Baker:2019gxo,Davies:2019irs,Paillas:2018wxs}.  The
RSD multipoles have the advantage of probing the velocity field and
the potential simultaneously, and the void-size dependence provides an
internal consistency check for chameleon screening.
The void size function~\cite{Sheth:2003py,Jennings:2013nsa,Pisani:2015jha,Contarini:2021fkv,Verza:2023rwl,Cautun:2017tkc}
is a particularly natural complement because it responds to the same
scale-dependent fifth force through the abundance of large and small
voids.
Complementary probes of modified gravity using redshift-space
higher-order statistics have also been investigated, including galaxy
bispectrum multipoles and post-reionization 21-cm bispectrum
multipoles in $\fR$ or related modified-gravity
scenarios~\cite{Pal:2025bisp,Pal:2026bisp21cm}.
Unlike these higher-order clustering probes, the present work focuses
on void--galaxy cross-correlation multipoles, which are particularly
sensitive to low-density environments where chameleon screening is
weakened.
Our void-based observable is also complementary to other projected
tests of $\fR$ gravity with future surveys, such as HI~21\,cm
intensity-mapping forecasts for BINGO and
SKA1-MID~\cite{Song:2026fr21cm}, since void multipoles probe the
coupled density, velocity, and potential fields around underdense
regions.

\textit{Limitations and future directions.}---Several
approximations in the present semi-analytical framework should be
kept in mind.
(i)~The universal void profile of Hamaus
\textit{et~al.}~\cite{Hamaus:2014afa} is used as the GR
template, with the $\fR$ modification applied in Fourier space
via the growth ratio $\mathcal{R}(k,a)$.  In $\fR$ simulations
the profile shape parameters ($\alpha$, $\beta$, $r_s/r_v$) may
themselves depend on the gravity
model~\cite{Cautun:2018gae,Falck:2017rvl}; accounting for this
shape variation could modify the predicted deviations at the
$\sim 10$--$20\%$ level.
(ii)~Galaxy bias is treated as a constant, with $b=2$ used as the
fiducial value.  Inside voids
the effective bias may differ from the mean value, and $\fR$ gravity
can introduce environment-dependent corrections to the
bias~\cite{Cautun:2018gae}.  The
quadrupole, being dominated by velocity terms, is largely immune to
this (Sec.~\ref{sec:results}); our explicit stability test over
$1.5\leq b\leq2.5$ changes the monopole MG deviation by
$\lesssim1.5\%$ of its peak and the quadrupole by
$\lesssim0.5\%$.  This supports the fiducial $b=2$ choice for the
semi-analytical forecast, but not as a replacement for a
survey-specific bias calibration.
Possible scale-dependent bias corrections in $\fR$ simulations would
need to be calibrated with mocks before turning the present
semi-analytical forecast into a precision constraint~\cite{Terasawa:2025png}.
(iii)~The spherical shell ODE uses a single effective wavenumber
$k_{\rm eff} = \pi/q$ per shell, neglecting mode mixing.  While the
amplification is robust to factor-of-4 variations in $k_{\rm eff}$
(Appendix~\ref{app:nl_keff}), coupling between different $k$-modes
could modify the NL amplification for small voids where $\Geff/G$
varies rapidly.
(iv)~The S/N estimates use a phenomenological synthetic covariance with
30\% profile scatter, a small scatter floor, and exponential radial
correlations inspired by void-correlation analyses;
non-Gaussian tails, survey masks, and super-sample variance could still
modify the effective S/N
(Appendix~\ref{app:covariance}).  The S/N values in
Tables~\ref{tab:sn_nl} and~\ref{tab:sn_direct}
should therefore be regarded as upper bounds.
(v)~Void exclusion and overlap: large voids may contain
sub-structures identified as small voids by watershed
algorithms~\cite{Sheth:2003py,Jennings:2013nsa}, and the
cross-covariance between overlapping void-size bins is not
captured by the independent-bin assumption in the current
multi-size bookkeeping.

Validation against $\fR$ $N$-body
simulations~\cite{Li:2011pj,Winther:2015wla,Brax:2012gr,Lombriser:2013wta}
is therefore the most important next step.  In particular, the
semi-analytical detectability estimates should be benchmarked against
mock catalogs with consistent void selection, survey masks, and
cross-bin covariance before being interpreted as precision
constraints.

The landscape of wide-field spectroscopic surveys has nevertheless
changed substantially since the original formalism was developed:
DESI now has public data products and an initial void
catalog~\cite{Rincon:2024desivast},
Euclid~\cite{Euclid:2011zbd} and Roman~\cite{Spergel:2015sza,Verza:2024roman}
provide the main Stage-IV high-redshift forecasts,
and Subaru PFS has entered science operations~\cite{PFS:2025ssp}.  Together these
surveys increase the available void statistics by roughly one to two
orders of magnitude relative to SDSS/BOSS-quality samples, making the
void-size dependent signal predicted here a realistic target.  The key
observable remains the transition from GR-like large voids to enhanced
small or intermediate voids, while multi-redshift measurements can map
$\lambda_C(z)$ and constrain $m_{\rm sc}^2(a)$.
The equivalence between configuration-space correlation functions and
harmonic-space power spectra as statistical estimators for
beyond-$\Lambda$CDM physics~\cite{Terasawa:2025png} is also relevant
for joint analyses combining void-RSD multipoles with photometric
and spectroscopic two-point statistics.

\textit{Void RSD as a precision probe of $\fR$ gravity.}---We
conclude that the void-galaxy cross-correlation RSD multipoles are
uniquely suited to constraining Hu-Sawicki $\fR$ gravity for
several reasons.  First, voids provide a low-density environment
where scalar screening is minimised, maximising the observable
MG signal.  Second, the void-size dependent transition from GR-like
to enhanced scalaron response directly maps the Compton wavelength
$\lambda_C$, providing a model-specific diagnostic that is
unavailable in cluster or galaxy clustering analyses.  Third, the
simultaneous access to the density field (monopole), velocity field
(quadrupole), and gravitational potential (dipole) allows for
internal consistency checks and breaks degeneracies with nuisance
parameters.  Fourth, the large NL amplification
($\mathcal{A}_0 \sim 4$--$10$) brings even weak signals into
the detectable regime.  With Stage-IV surveys (DESI, Euclid,
Roman) and complementary programs (Subaru PFS) now delivering data,
the theoretical framework developed in
this paper provides the foundation for turning these observations
into competitive constraints on $\fRz$.

\begin{acknowledgments}
Y.N. thanks Masahiro~Takada, Ryo~Terasawa, Joaquin~Armijo, Xin~Ren,
HongSheng~Zhao, Dongdong~Zhang, Elisa~G.~M.~Ferreira, Misao~Sasaki,
Sunao~Sugiyama, Yuuki~Sugiyama, and Kazushige~Ueda for useful
discussions and suggestions during the development of this work.
This work was initiated at the Kavli IPMU (WPI), the University of Tokyo,
where part of it was also carried out. This work was also partially performed at 
the Center for Data-Driven Discovery, Kavli IPMU (WPI). 
Y.N. especially thanks Kazuhiro~Yamamoto for the early-stage discussions that inspired this work.

This work was supported by the Japan Society for the Promotion of
Science (JSPS) KAKENHI Grant Number JP24K17041.
\end{acknowledgments}

\appendix

\section{Semi-Analytical Calculation Details}
\label{app:algorithm}

The nonlinear shell evolution procedure has three steps:
\begin{enumerate}
\item \textit{Backscaling.}---The observed void template
  $\delta_{\rm template}(r)$ at the target redshift $z_{\rm target}$
  is scaled back to $z_{\rm init} = 50$ using the linear growth
  factor:
  $\delta_{\rm init}(q) = \delta_{\rm template}(q)\;
  D(z_{\rm init}) / D(z_{\rm target})$,
  where $q \approx r$ in the linear regime.  Initial conditions for
  the ODE are set to the growing mode:
  $d\delta/dz|_{z_{\rm init}} = -\delta_{\rm init}/(1+z_{\rm init})$.

\item \textit{Shell evolution.}---Each of the $N_{\rm shell} = 200$
  concentric shells is evolved from $z_{\rm init}$ to $z_{\rm target}$
  by integrating Eq.~\eqref{eq:nl_ode}, once with $\Geff/G = 1$
  (GR) and once with $\Geff(k_{\rm eff}(q),a)/G$ ($\fR$).
  The integration is also performed with the linear
  Eq.~\eqref{eq:lin_ode} to obtain the linear predictions for
  comparison.

\item \textit{Multipole computation.}---The nonlinear density profile
  $\delta_{\rm NL}(r)$ replaces the linear $\delta(r)$ in the
  calculation of Secs.~\ref{sec:velocity}--\ref{sec:rsd}.  The derived
  quantities $\bar{\Delta}(r)$, $\tilde{V}(r)$, and $\psi(r)$ are
  recomputed from $\delta_{\rm NL}$, and the RSD multipoles
  $\xi_\ell^{\rm NL}(s)$ are evaluated using the full formulas of
  Sec.~\ref{sec:rsd}.
\end{enumerate}
At small $r$ the mean density
$\bar{\Delta}(r)$ is regularized via Taylor expansion around
$r = 0$: $\bar{\Delta} \approx \delta(0) + \delta''(0)\,r^2/10$
for $r < 0.5\;\Mpc$.  Numerical resolution and integrator settings are
reported separately in the technical documentation.

\section{Nonlinear Amplification Robustness}
\label{app:nl_keff}

The spherical shell ODE Eq.~\eqref{eq:nl_ode} requires an effective wavenumber
$k_{\rm eff}$ to evaluate $\Geff(k_{\rm eff}, a)$ for each shell at
Lagrangian radius $q$.  We adopt $k_{\rm eff} = C \cdot \pi/q$ with
$C = 1$ as the fiducial choice.  Varying $C \in \{0.5,\,0.75,\,1.0,\,
1.5,\,2.0\}$ (Table~\ref{tab:keff}), the shell-level NL amplification $\mathcal{A}_0$
varies by only $\sim 6\%$ across a factor-of-4 range in $k_{\rm eff}$
(from $3.1$ to $3.3$).
The fractional monopole deviation $\Delta\xi_0/\xi_0^{\rm GR}$
varies by a factor of $\sim 2.4$ (from $+0.016$ to $+0.038$),
reflecting the change in $\Geff/G$ at different $k$, rather than
a breakdown of the shell approximation.

\begin{table}[tb]
\caption{$k_{\rm eff}$ sensitivity for $\fR$ ($\fRz = 10^{-5}$,
$r_v = 30\;\Mpc$): fractional MG deviation $\Delta\xi_0/\xi_0^{\rm GR}$
at the monopole peak and NL amplification factor $\mathcal{A}_0$
as a function of $C$.}
\label{tab:keff}
\begin{ruledtabular}
\begin{tabular}{ccccc}
$C$ & $k_{\rm eff}$ [$h/$Mpc] & $\Geff/G$
  & $\Delta\xi_0/\xi_0^{\rm GR}$ & $\mathcal{A}_0$ \\
\hline
0.50 & 0.052 & 1.016 & $+0.016$ & $3.1$ \\
0.75 & 0.079 & 1.025 & $+0.024$ & $3.2$ \\
1.00 & 0.105 & 1.031 & $+0.028$ & $3.2$ \\
1.50 & 0.157 & 1.038 & $+0.034$ & $3.3$ \\
2.00 & 0.209 & 1.041 & $+0.038$ & $3.3$ \\
\end{tabular}
\end{ruledtabular}
\end{table}

Table~\ref{tab:nl_all} extends the observable-level amplification to
all three void size classes.  The increase of $\mathcal{A}_0$ for
smaller voids reflects deeper profiles and stronger $\Geff/G$; the
dipole amplification $\mathcal{A}_1 \approx 3.5$--$4.3$ is
roughly constant because the potential contribution involves a
volume average that moderates the deep-void NL effects.

\begin{table}[tb]
\caption{Observable-level NL amplification $\mathcal{A}_\ell$ for
all void sizes ($\fRz = 10^{-5}$, $z = 0.5$), defined via
Eq.~\eqref{eq:Amp_def}.}
\label{tab:nl_all}
\begin{ruledtabular}
\begin{tabular}{lccccc}
Void class & $r_v$ [$\Mpc$] & $\Geff/G$ & $\mathcal{A}_0$ & $\mathcal{A}_1$ & $\mathcal{A}_2$ \\
\hline
Large  & 30.0 & 1.031 & $3.7$ & $4.2$ & $4.3$ \\
Medium & 17.6 & 1.063 & $5.8$ & $3.6$ & $3.7$ \\
Small  & 11.7 & 1.131 & $10.2$ & $3.5$ & $5.8$ \\
\end{tabular}
\end{ruledtabular}
\end{table}

\section{Void+Filament Model}
\label{app:filament}

The right panel of Fig.~\ref{fig:potential} shows the density
profile of a void embedded in a filamentary environment.  In the
cosmic web, voids are bounded by walls, filaments, and nodes that
produce an overdense ridge beyond the compensation wall of the
Hamaus profile~\cite{Hamaus:2014afa,Sheth:2003py}.

For an isolated void, mass conservation requires compensation:
$\int_0^\infty \delta(r)\,r^2\,dr = 0$.
\label{eq:compensation}
We model the excess overdensity from the cosmic web with an
additive Gaussian ridge:
\begin{equation}
\delta_{\rm fil}(r) = A_{\rm fil}\,\exp\!\Bigl[
  -\frac{(r - r_{\rm fil})^2}{2\sigma_{\rm fil}^2}\Bigr]\,,
\label{eq:filament}
\end{equation}
with $r_{\rm fil} = 1.5\,r_v$ (mean void-filament
separation~\cite{Hamaus:2014afa,Cautun:2013lga}), $\sigma_{\rm fil}
= 0.2\,r_v$ (typical filament cross-section $4$--$6\;\Mpc$ in
simulations~\cite{Cautun:2013lga}), and $A_{\rm fil} = 0.1$
(from mass conservation: $A_{\rm fil} \approx 0.08$--$0.12$ for
the large void class).  The filament ridge lies at $r \sim 1.5\,r_v
\gg \lambda_C$, where screening ensures $\Geff/G \to 1$, so its
primary effect is to modify the compensation region of the potential
at the $\lesssim 1\%$ level without affecting the MG signal.

\section{Survey Specifications and Covariance Model}
\label{app:covariance}

\subsection{Survey parameters}

Table~\ref{tab:surveys} lists the surveys considered.  The approximate
void counts $N_v$ are baseline usable-count estimates for stacked
spectroscopic RSD analyses, anchored to published BOSS catalogs,
Euclid Flagship mock forecasts, DESI data releases, and a
volume-scaled PFS estimate
\cite{Hamaus:2015gua,Hamaus:2021lzy,DESI:2016fyo,DESI:2025dr1,DESI:2025dr2bao,Takada:2014pfs,Tamura:2016wso,Tamura:2024spie,Tamura:2025natastro,PFS:2025ssp}.
For DESI~Y5 and Subaru~PFS, where no directly matching high-redshift
RSD-ready void catalog is yet available, we estimate $N_v$ by volume
scaling from BOSS:
\begin{equation}\label{eq:Nv_scaling}
  N_{v}^{\rm est} \;\approx\;
  \rho_v^{\rm BOSS}\,V_{\rm survey}\,
  \Bigl(\frac{\bar{n}_g}{\bar{n}_g^{\rm BOSS}}\Bigr)^{1/2},
\end{equation}
where $V_{\rm survey}$ is the comoving survey volume and the
$\bar{n}_g^{1/2}$ factor approximately accounts for the fact that
denser tracer samples resolve more small voids.  This gives a
conservative DESI~Y5 baseline of $N_v\sim 4\times10^4$ and a PFS
baseline of $N_v\sim 1.2\times10^4$.  The ``Combined'' entry is only a
bookkeeping sum for the compressed template S/N estimates, not a
standalone survey specification.  All such values should be regarded
as order-of-magnitude guides with $\sim 30$--$50\%$ systematic
uncertainty from void-finder choice, minimum radius, and tracer
selection.

\begin{table}[tb]
\caption{Survey parameters adopted for the S/N estimates.
$\bar{n}_g$ is the effective tracer number density in units of
$(h/\mathrm{Mpc})^3$ and $N_v$ the approximate usable void count
after basic quality cuts.  BOSS is quoted for the DR12
LOWZ+CMASS footprint~\cite{Hamaus:2020rpa}; Euclid follows the Flagship mock scale of
Ref.~\cite{Hamaus:2021lzy}; DESI uses ELG-dominated tracer densities
relevant for $0.6<z<1.6$~\cite{Rincon:2024desivast,DESI:2026milestone}; Subaru PFS uses its planned [O\,\textsc{ii}]
survey sample; Roman refers to the reference-HLSS void
forecast~\cite{Spergel:2015sza,Greene:2019jlm,Wang:2022roman,Verza:2024roman}; Euclid follows~\cite{Euclid:2011zbd};
LSST~\cite{LSST:2008ijt} is listed only as a photometric reference.  The
Combined row is a bookkeeping aggregate for template S/N estimates,
not a standalone survey specification.}
\label{tab:surveys}
\begin{ruledtabular}
\begin{tabular}{lcccc}
Survey & Area [deg$^2$] & $z$ range & $\bar{n}_g$ & $N_v$ \\
\hline
BOSS        & 9\,376  & 0.15--0.70  & $3{\times}10^{-4}$ & $5{\times}10^3$ \\
DESI Y5 (ELG)& 14\,000  & 0.6--1.6 & $5{\times}10^{-4}$ & $4{\times}10^4$ \\
Euclid      & 14\,000  & 0.9--1.8  & $3{\times}10^{-4}$ & $4.4{\times}10^4$ \\
Subaru PFS  &  1\,400  & 0.8--2.4  & $4{\times}10^{-4}$ & $1.2{\times}10^4$ \\
Roman ref.\ HLSS &  2\,000  & 1.0--3.0  & $\sim10^{-3}$      & ${>}8{\times}10^4$ \\
LSST (phot.)& 18\,000  & broad      & ---                & --- \\
Combined    & $\sim$30\,000 & 0.6--2.4 & $5{\times}10^{-4}$ & $10^5$ \\
\end{tabular}
\end{ruledtabular}
\end{table}

\subsection{Covariance model}

The analyses of
Refs.~\cite{Hamaus:2015gua,Hamaus:2016sei,Hamaus:2021lzy}
estimate
the covariance of the stacked void--galaxy correlation from
jackknife or mock-catalog realizations and use the resulting full
covariance matrix in the likelihood.  In the absence of such
modified-gravity mock catalogs, we use a \emph{synthetic}
or phenomenological covariance that mimics the non-diagonal
structure of a full covariance matrix, bearing in mind we are not 
claiming a full Fisher forecast but rather a controlled sensitivity test of the S/N scaling with $N_v$ under a reasonable covariance structure.  The synthetic covariance is
inspired by the observed covariance structure of void correlation functions in
simulations and data~\cite{Hamaus:2015gua,Hamaus:2016sei,Hamaus:2021lzy,Nadathur:2019mct,Woodfinden:2022bhx} and is designed to capture the key features that affect the S/N scaling: the diagonal amplitude (which sets the overall noise level) and the off-diagonal correlations (which affect how information from different radial bins and multipoles combines).  It is not intended to capture all the detailed features of a survey-specific covariance matrix, which would require mock catalogs with the same void-finding algorithm, tracer selection, and survey geometry.
For the combined data-vector index $a=(\ell,i)$, where $i$ labels radial bins and $\ell$ labels
multipoles, we take
\begin{equation}
C^{\rm syn}_{(\ell i),(\ell' j)}
  = \sigma_{\ell i}\sigma_{\ell' j}\,
    \rho_{\ell\ell'}\,
    \exp\!\left[-\frac{|s_i-s_j|}{L_{\rm corr}}\right] ,
\label{eq:cov_syn}
\end{equation}
with diagonal amplitude
\begin{equation}
\begin{split}
\sigma_{\ell i}^2 &=
  \frac{2\ell+1}{N_v\,\bar{n}_g\,V_{{\rm shell},i}}
  + \frac{[f_{\rm rel}\xi_\ell^{\rm GR}(s_i)]^2
  + \sigma_{\ell,{\rm floor}}^2}{N_v}\,,\\
V_{{\rm shell},i} &= 4\pi s_i^2\Delta s .
\end{split}
\label{eq:var_syn}
\end{equation}
The first term is the Poisson pair-count contribution and the second
term models void-to-void profile scatter.  We use
$f_{\rm rel}=0.30$, $L_{\rm corr}=0.5R_v$, and
$\sigma_{\ell,{\rm floor}}=f_{\rm floor}\max_i|\xi_\ell^{\rm GR}(s_i)|$
with $f_{\rm floor}=0.03$ to prevent artificial error zeros at
multipole zero crossings.  The default cross-multipole correlations
are $\rho_{\ell\ell}=1$, $\rho_{02}=0.20$, and
$\rho_{01}=\rho_{12}=0.10$.  This covariance enters
Eq.~\eqref{eq:sn_cum} through the full inverse matrix.  It is a
controlled sensitivity test, not a substitute for the survey-specific
full covariance matrices advocated and used in the Hamaus/Euclid
void analyses.  The choice $L_{\rm corr}=0.5R_v$ is a fiducial
void-scale correlation length: it correlates neighboring radial bins
within the coherent stacked profile while still allowing broad
shape information to contribute.  It should be varied in future
mock-calibrated forecasts rather than interpreted as a measured
covariance length.

Effects such as non-Gaussian profile scatter, survey-specific
geometry, void-finder dependence, and void exclusion/overlap can still
shift the effective S/N, so the compressed-template estimates of
Sec.~\ref{sec:detectability} should be interpreted as sensitivity
tests rather than full likelihood forecasts.

\begin{figure*}[tb]
\includegraphics[width=\textwidth]{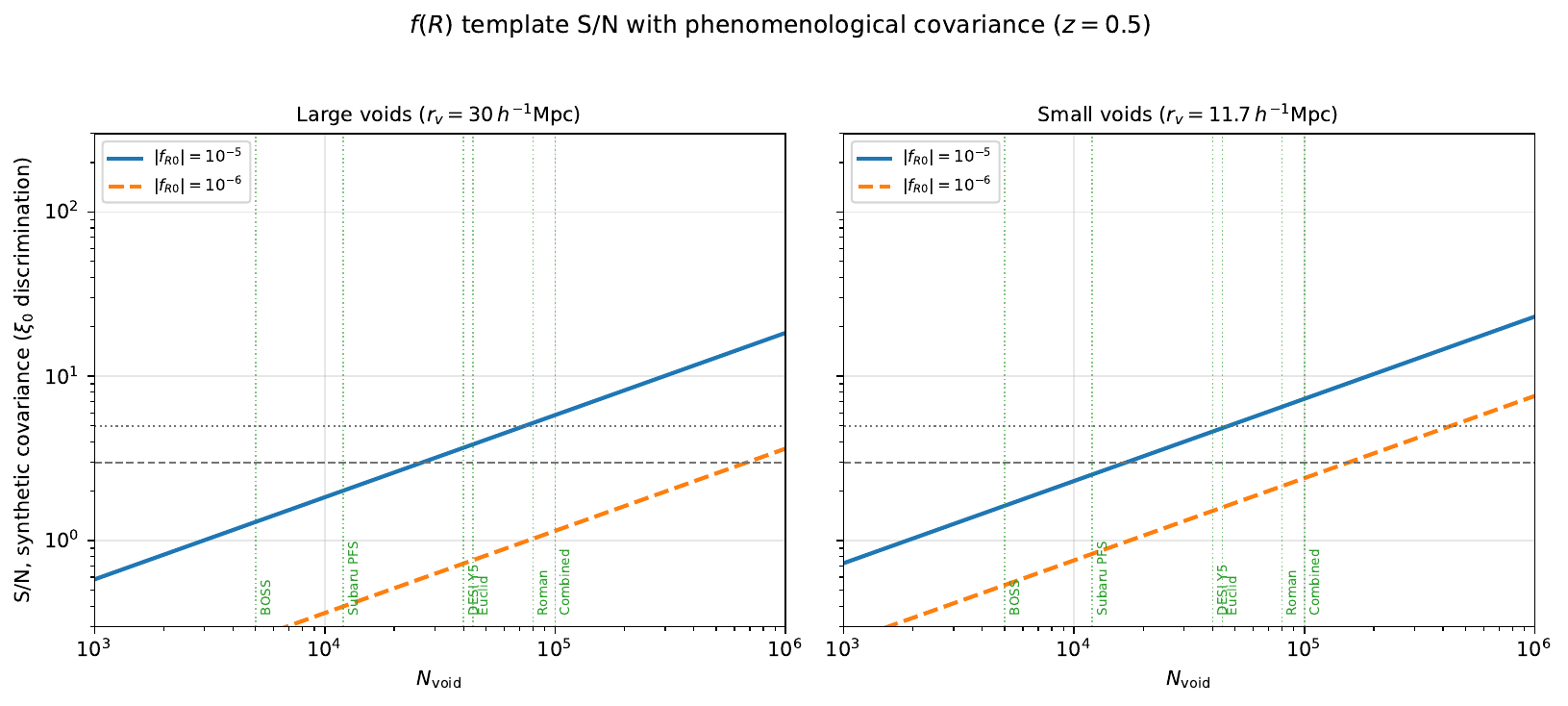}
\caption{Synthetic-covariance template-Fisher S/N for the
$\fR$-vs-GR monopole discrimination as a function of the number of
stacked voids at $z=0.5$.  Left: large voids
($r_v = 30\;h^{-1}$Mpc); right: small voids
($r_v = 11.7\;h^{-1}$Mpc).  Solid: $\fRz = 10^{-5}$;
dashed: $\fRz = 10^{-6}$.  Green vertical markers indicate current
and planned survey void counts (Table~\ref{tab:surveys}), including
the Roman reference HLSS forecast.  Horizontal lines mark the
$3\sigma$ and $5\sigma$ thresholds.  This figure is a diagnostic of
the synthetic covariance model and should not be interpreted as a
full survey likelihood forecast.}
\label{fig:fisher_sn_diag}
\end{figure*}

The same synthetic covariance can be used to draw diagnostic
monopole S/N-versus-$N_v$ curves.  Figure~\ref{fig:fisher_sn_diag}
shows this dependence for the large- and small-void templates; it
preserves the same hierarchy as Table~\ref{tab:sn_nl}, with small
voids reaching higher S/N at fixed $N_v$, but is not a substitute for
a mock-calibrated forecast.

\bibliography{references_fr}

\end{document}